\documentclass[draftclsnofoot,onecolumn,12pt]{IEEEtran}
\usepackage{graphicx}
\usepackage{amsmath}
\usepackage{amssymb}
\usepackage{subfigure}
\usepackage{verbatim}
\usepackage{fixltx2e}

\usepackage{xcolor}
\usepackage[final,ulem=normalem]{changes}
\definechangesauthor{RW}{blue}
\definechangesauthor{TC}{red}
\definechangesauthor{BX}{green}
\long\def\comment#1{}

\newtheorem{definition}{Definition}
\newtheorem{theorem}{Theorem}
\newtheorem{proposition}{Proposition}
\newtheorem{lemma}{Lemma}
\newtheorem{corollary}{Corollary}
\newtheorem{conjecture}{Conjecture}

\begin{document}
\date{}
\title{Optimal Encoding Schemes for Several Classes of Discrete Degraded Broadcast Channels}
\author{\normalsize{Bike Xie, {\it Student Member,
IEEE}, Thomas Courtade, {\it Student Member, IEEE}, \\ and Richard D.
Wesel, {\it Senior Member, IEEE}}
\thanks{This work was supported by the Defence Advanced Research
Project Agency SPAWAR Systems Center, San Diego, California under
Grant N66001-02-1-8938.
\newline This work was done when Bike Xie was with the Electrical
Engineering Department,University of California, Los Angeles. Bike
Xie is currently with Marvell Semiconductor Inc.
Thomas Courtade and Richard Wesel are
with the Electrical Engineering Department,University of California,
Los Angeles, CA 90095 USA. (e-mail:
bike@marvell.com;
tacourta@ucla.edu; wesel@ee.ucla.edu)}}

\maketitle
\begin {abstract}
Consider a memoryless degraded broadcast channel (DBC) in which the
channel output is a single-letter function of the channel input and
the channel noise.  As examples, for the Gaussian broadcast channel
(BC) this single\added[BX]{-}letter function is regular Euclidian
addition and for the binary-symmetric BC this \added{single-letter}
function is Galois-Field-two addition.  This paper identifies
several classes of discrete memoryless DBCs for which a relatively
simple encoding scheme, which we call natural encoding, achieves
capacity.   Natural Encoding (NE) combines symbols from independent
codebooks (one for each receiver) using  the same single-letter
function that adds distortion to the channel.  The alphabet size of
each NE codebook is bounded by that of the channel input.

\added{Inspired by Witsenhausen and Wyner, this paper defines the conditional entropy bound
function $F^*$, studies its properties, and applies them to show
that NE achieves the boundary of the capacity region for the
multi-receiver broadcast Z channel.  Then, this paper defines the input-symmetric DBC, introduces permutation
encoding for the input-symmetric DBC,  and proves its
optimality.  Because it is a special case of permutation encoding, NE is capacity achieving for the two-receiver
group-operation DBC.  Combining the broadcast Z channel and group-operation DBC results yields a proof that NE is also optimal for the discrete multiplication DBC.  Along the way, the paper also provides explicit parametric expressions for the two-receiver binary-symmetric DBC and broadcast Z channel.}

\end {abstract}

\begin {keywords}
Degraded broadcast channel, natural encoding, broadcast Z channel,
input-symmetric, group-\replaced{operation}{additive} degraded broadcast channel, discrete
multiplication degraded broadcast channel, Gaussian broadcast channel, binary-symmetric broadcast channel.
\end {keywords}

\section{Introduction}
\subsection{Background}
\replaced{Nearly four decades ago}{In the 70's}, Cover \cite{Cover1972}, Bergmans \cite{Bergmans1973}
and Gallager \cite{Gallager1974} established the capacity region for
degraded broadcast channels (DBC). A common optimal transmission
strategy to achieve the boundary of the capacity region for DBCs is
the joint encoding scheme presented in \cite{Cover1972}
\cite{Bergmans1973}. Specifically, the information intended for the receiver
with the most degraded channel is encoded to produce a first codeword.   \replaced{Conditioned on that first codeword, a codebook is selected for the receiver with the second most degraded channel}{Selection of an appropriate codebook for the receiver with the second most degraded channel depends on that first codeword},
and so forth.

There is at least one independent-encoding scheme (in which the codebook for each user is independent of the messages intended for other users) that can  achieve the capacity of any
DBC \cite{TelatarPrivate}. This scheme essentially embeds all
symbols from all the needed codebooks for the less-degraded receiver(s)
into a single super-symbol (but perhaps with a large alphabet). Then
a single-letter function uses the input symbol from the
more-degraded receiver to extract the needed symbol from the super
symbol provided by the less-degraded receiver. See Appendix \ref{app:A} for a
detailed description of this encoding scheme.

Cover \cite{Cover1975} introduced an independent-encoding scheme for
two-receiver broadcast channels (BCs). When applied to two-receiver
DBCs, this scheme independently encodes receivers' messages, and
then combines these resulting codewords by applying a single-letter
function. This scheme does not specify what codebooks to use or what
single-letter function to use. It is a general independent-encoding
approach, which includes the independent-encoding scheme described in Appendix
\ref{app:A}.

Consider DBCs in which the received signal of each component channel
can be modeled as a single-letter function of the channel input and
the channel noise. A simple encoding scheme that is optimal for some
of those DBCs is an independent-encoding approach in which symbols
from independent codebooks, each with the same alphabet as the
channel input, are combined using the same single-letter function
that adds distortion to the channel. We refer to this encoding
scheme as the natural encoding (NE) scheme. As an example, the NE
scheme for a two-receiver Gaussian BC has as each transmitted symbol
the real addition of two real symbols from independent codebooks.
The NE scheme is known to achieve the boundary of the capacity
region for several BCs including Gaussian BCs \cite{Bergmans1974},
binary-symmetric BCs \cite{Bergmans1973} \cite{WynerPartII}
\cite{Witsenhausen1974} \cite{Witsenhausen1975}, discrete additive
DBCs \cite{Benzel1979} and two-receiver broadcast Z channels
\cite{BZC} \cite{Bike2007}.

In proving the optimality of NE schemes for Gaussian BCs
and binary-symmetric BCs, Shannon's entropy power inequality (EPI) \cite{shannon1948} and
``Mrs. Gerber's Lemma'' \cite{WynerPartI}, respectively, play the same significant
role.
Shannon's EPI gives a lower
bound on the differential entropy of the sum of independent random
variables. In Bergmans's remarkable paper \cite{Bergmans1974}, he
applied the EPI to establish a converse showing the optimality of
the scheme given by \cite{Cover1972} \cite{Bergmans1973} (the NE
scheme) for Gaussian BCs. Similarly, ``Mrs. Gerber's Lemma''
provides a lower bound on the entropy of a sequence of
binary-symmetric channel outputs. Wyner and Ziv obtained ``Mrs.
Gerber's Lemma'' and applied it to establish a converse showing that
the NE scheme for binary-symmetric BCs suggested by Cover
\cite{Cover1972} and Bergmans \cite{Bergmans1973} achieves the
boundary of the capacity region \cite{WynerPartII}.

Witsenhausen and Wyner made two seminal contributions in
\cite{Witsenhausen1974} and \cite{Witsenhausen1975}: the notion of
minimizing one entropy under the constraint that another related
entropy is fixed, called the conditional entropy bound, and the use
of input symmetry as a way of solving an entire class of channels
with a single unifying approach. Witsenhausen and Wyner applied the
first idea to establish an outer bound of the capacity region for
DBCs \cite{Witsenhausen1975}. For binary-symmetric BCs, this outer
bound coincides with the capacity region, which proved once more
that the NE scheme for binary-symmetric BCs is capacity-achieving.

Later, Benzel \cite{Benzel1979} applied the conditional entropy
bound to prove that the capacity regions for discrete additive
degraded interference channels (DADICs) and the corresponding
discrete additive DBC are the same, which means that NE
is capacity-achieving for discrete additive DBCs. Recently Liu and
Ulukus \cite{DDIC} \cite{Liu2008} extended Benzel's results to
include the larger class of discrete degraded interference channels
(DDICs). For these DDICs, Liu and Ulukus introduced a
capacity-achieving independent encoding scheme for the
corresponding DBCs as long as the transmitted signal for the DBC can
be appropriately defined.

\subsection{\replaced{Contributions}{Organization}}
\replaced{
The main contributions of this paper are the following:
\begin{enumerate}
\item Establishing that NE is capacity-achieving for multi-receiver broadcast Z
channels
\item Introducing permutation encoding for
input-symmetric DBCs and proving its optimality
\item Proving the
optimality of the NE scheme for discrete multiplication DBCs.
\end{enumerate}
}
{The main contributions of this paper are: (1) establishing that NE is capacity-achieving for multi-receiver broadcast Z
channels; (2) introducing permutation encoding for
input-symmetric DBCs and proving its optimality; (3) proving the
optimality of the NE scheme for discrete multiplication DBCs.}

This paper begins its investigation by extending ideas from
Witsenhausen and Wyner \cite{Witsenhausen1975} to study a
conditional entropy bound for the channel output of a discrete DBC.
This conditional entropy bound leads to a representation of the
capacity region of discrete DBCs. \added{As an application, explicit
parametric expressions for the capacity regions are derived for
two-}\replaced[BX]{receiver}{user}\added{ binary-symmetric BCs and
two-}\replaced[BX]{receiver}{user}\replaced{ broadcast Z
channels.}{This paper simplifies the expression of the conditional
entropy bound for binary-symmetric BCs and broadcast Z channels.}
For broadcast Z channels, this simplified expression of the
conditional entropy bound demonstrates that the NE scheme identified
as optimal for two-receiver broadcast Z channels in \cite{BZC} is
also optimal for more than two receivers.

This paper then defines what it means for a degraded broadcast
channel to be input-symmetric (IS) (first introduced in
\cite{Witsenhausen1975} for point-to-point channels) and provides an
independent-encoding scheme, referred to as permutation encoding,
which achieves the capacity region of all IS-DBCs. The
group-\replaced{operation}{additive} DBC, which includes the
discrete additive DBC \cite{Benzel1979} as a special case, is a
class of input-symmetric DBCs for which each channel output is a
group \replaced{operation}{addition}\footnote{A group
\replaced{operation}{addition} is an operation which satisfies the
group axioms (Closure, Associativity, Identity element, Inverse
element) on a pre-defined set. The group \deleted{addition}
operation and the set together forms a group.} of the channel input
and the channel noise.  For group-\replaced{operation}{additive}
DBCs, permutation encoding is equivalent to NE, establishing the
optimality of NE for group-\replaced{operation}{additive} DBCs.

The discrete multiplication DBC is a discrete DBC for which each channel
output is a discrete multiplication\footnote{The definition of the
discrete multiplication is given in Section
\ref{sec:multiplication}. We refer to this operation as discrete
multiplication because it is a generalization of multiplication
as defined in a field.} of the channel input and the channel noise. This paper concludes its investigations by applying the conditional entropy bound to discrete multiplication DBCs and proving that NE
achieves the boundary of the capacity region in this case.

\subsection{\added{Organization}}

This paper is organized as follows: Subsection
\ref{sec:notation} below lays out the notation used in this paper.
Section
\ref{sec:Fstar} defines and studies the conditional entropy bound
$F^*(\boldsymbol{q},s)$ for the channel output of a discrete DBC,
and represents the capacity region of the discrete DBC using the
function $F^*(\boldsymbol{q},s)$. Section \ref{sec:evaluation} uses
duality to evaluate $F^*(\boldsymbol{q},s)$ and provides
an approach to characterizing optimal transmission
strategies for the discrete DBC based on this evaluation.  As
an example, Section \ref{sec:BSC} uses the duality-based computation
of  $F^*(\boldsymbol{q},s)$ to provide an explicit
parametric expression for the capacity
region of the two-receiver
binary-symmetric BC. Section \ref{sec:BZC} proves
the optimality of the NE scheme for broadcast Z channels with more
than two receivers.
Section \ref{sec:inputsymmetric} defines the IS-DBC,
introduces the permutation encoding approach, and proves its
optimality for IS-DBCs.  Section \ref{sec:multiplication} studies
the discrete multiplication DBC and shows that NE achieves the
boundary of the capacity region for the discrete multiplication DBC.
Section \ref{sec:conclusion} delivers the conclusions.

\subsection{Notation} \label{sec:notation}
Denote $X \rightarrow Y$ as a discrete memoryless channel with
channel input $X$ and output $Y$. Denote $X \rightarrow
Y^{(1)} \rightarrow \cdots \rightarrow Y^{(K)}$ as a $K$-receiver
($K \geq 2$) discrete memoryless DBC where $X$ is the channel input,
and $Y^{(i)}$ ($i=1,\cdots,K$) is the $i$-th least-degraded
output. For simplicity of notation, we also denote $X \rightarrow Y
\rightarrow Z$ as a two-receiver DBC where $Y$ is the less-degraded
output and $Z$ is the more-degraded output. Since the capacity region of a
statistically-degraded BC without feedback is equivalent to that of the
corresponding physically-degraded BC with the same marginal transition
probabilities, we assume the DBCs in this paper are physically
degraded without loss of generality. Hence, $X \rightarrow Y
\rightarrow Z$ also denotes a Markov chain, i.e.,
$\text{Pr}(Z=z|Y=y,X=x) = \text{Pr}(Z=z|Y=y)$.

Throughout this paper, we use $X$ to represent a scalar random
variable at the channel input. Denote $x$ and $\mathcal{X}$ as its
specific value and its alphabet respectively. We also denote
$\boldsymbol{X}$ as a sequence of random variables of length $N$ at
the channel input. $\boldsymbol{x}$ denotes its specific value.
$X_i$ and $x_i$ denote the $i$-th element of $\boldsymbol{X}$ and
$\boldsymbol{x}$ respectively. We apply the same notation rules to
the channel outputs $Y$, $Z$, $Y^{(i)}$, the auxiliary random
variable $U$, and the codeword $\boldsymbol{X}^{(i)}$ for the $i$-th receiver.

Let $X \rightarrow Y \rightarrow Z$ be a two-receiver discrete
memoryless DBC where $\mathcal{X}$ $=\{1,2,\cdots,k\}$,
$\mathcal{Y}$ $= \{1,2,\cdots,n\}$, and $\mathcal{Z}$ $=
\{1,2,\cdots,m\}$. Let $T_{YX} $ be an $n \times k$ stochastic
matrix with entries $T_{YX}(j,i)=\text{Pr}(Y=j|X=i)$ and $T_{ZX} $
be an $m \times k$ stochastic matrix with entries $T_{ZX}
(j,i)=\text{Pr}(Z=j|X=i)$. Thus, $T_{YX} $ and $T_{ZX}$ are the
marginal transition probability matrices of the degraded broadcast
channel.

In this paper, we denote column vectors $\boldsymbol{p}$,
$\boldsymbol{q}$, and $\boldsymbol{w}$ as the distributions of
discrete random variables. In particular, $\boldsymbol{p}_{X}$
denotes the distribution of $X$. Let $\Delta_{n}
= \big\{ (p_1,\cdots,p_n) \in \mathbb{R}$ $| \sum_{i=1}^{n}p_n = 1,
\text{and } p_i \geq 0 \text{ for all } i \big\}$
denote the unit $(n-1)$-simplex of probability
$n$-vectors. We denote $h_{n}: \Delta_{n} \mapsto \mathbb{R}$ as the
entropy function for $n \geq 2$, i.e., $h_{n}([p_1,\cdots ,p_n]^{T})
\triangleq h_{n}(p_1,\cdots ,p_n)\triangleq -\sum p_i\ln p_i$. We
also denote $h: [0,1] \mapsto \mathbb{R}$ as $h(p) \triangleq
h_2([p,1-p]^{T})$.

Following the traditional notation, we denote $H(X)$ as the entropy
of $X$, $H(Y|X)$ as the conditional entropy of $Y$ given $X$,
$I(X;Y)$ as the mutual information between $X$ and $Y$, and
$I(X;Y|U)$ as the mutual information between $X$ and $Y$ given $U$.  \added{Since we have defined $h_n(\cdot)$ using the natural logarithm, all information quantities considered in this paper are in terms of \emph{nats}, unless explicitly stated otherwise.}

\section{The Conditional Entropy Bound $F^*(\boldsymbol{q},s)$}
\label{sec:Fstar}
\added{Observe that any auxiliary random variable $U$ with alphabet size $l \geq 1$ is
characterized by its distribution $\boldsymbol{w} =
[w_1,\cdots,w_l]^T \in \Delta_{l}$ and the transition
probability matrix from $U$ to $X$, $T_{XU}=
[\boldsymbol{t}_1 \cdots \boldsymbol{t}_l]$ where
$\boldsymbol{t}_{j} \in
\Delta_{k}$ for $j = 1,
\cdots, l$.}
The following definition introduces a conditional entropy bound central to our analysis:
\begin{definition} ($F^*_{T_{YX},T_{ZX}}(\boldsymbol{q},s)$)
Let \replaced{$\boldsymbol{q}\in \Delta_{k}$}{vector $\boldsymbol{q}$ in the simplex $\Delta_{k}$} be the
distribution of the channel input $X$. \replaced{The function
$F^*_{T_{YX},T_{ZX}}(\boldsymbol{q},s)$ is defined as
\begin{align}
F^*_{T_{YX},T_{ZX}}(\boldsymbol{q},s) = \inf_{\substack{
p(u,x)~:~H(Y|U)=s,~\boldsymbol{p}_{X} = \boldsymbol{q},\\
\mbox{\begin{scriptsize}~and~\end{scriptsize}} U \rightarrow X
\rightarrow (Y,Z)}} H(Z|U).
\end{align}
}{The function
$F^*_{T_{YX},T_{ZX}}(\boldsymbol{q},s)$ is the infimum of $H(Z|U)$
with respect to all discrete random variables $U$ such that a) $H(Y|U) = s$;
 and b) $U$ and $Y,Z$ are conditionally independent
given $X$, i.e., $U \rightarrow X \rightarrow Y,Z$ forms a Markov
chain.}
\end{definition}

Thus $F^*(\boldsymbol{q},s)$ is essentially the smallest possible value of  $H(Z|U)$ given a specified input distribution and a specified value of $H(Y|U)$.  We will sometimes abbreviate $F^*_{T_{YX},T_{ZX}}(\boldsymbol{q},s)$ to $F^*(\boldsymbol{q},s)$ or even $F^*(s)$ when there is sufficient context to avoid confusion.

\deleted{Any auxiliary random variable $U$ with alphabet size $l
\geq 1$ is characterized by its distribution $\boldsymbol{w} =
[w_1,\cdots,w_l]^T \in \Delta_{l}$ and the transition probability
matrix from $U$ to $X$, $T_{XU}= [\boldsymbol{t}_1 \cdots
\boldsymbol{t}_l]$ where $\boldsymbol{t}_{j} \in \Delta_{k}$ for $j
= 1, \cdots, l$.} The choices of $p(u,x)$ satisfying the conditions
\added{$H(Y|U)=s$, $\boldsymbol{p}_{X} = \boldsymbol{q}$, and $U
\rightarrow X \rightarrow (Y,Z)$} in the definition of
$F^*_{T_{YX},T_{ZX}}(\boldsymbol{q},s)$ correspond to the choices of
$l$,$\boldsymbol{w}$ and $T_{XU}$ such that
\begin{equation}
\boldsymbol{q} = \boldsymbol{p}_{X} = T_{XU} \boldsymbol{w}=\sum _{j
=1}^{l}w_{j} \boldsymbol{t}_{j} \label{eq:p}
\end{equation}
and
\begin{equation} s  = H(Y|U) = \sum _{j =1}^{l}w_{j}
h_{n}(T_{YX} \boldsymbol{t}_{j}). \label{eq:xi}
\end{equation}
The corresponding $H(Z|U)$ \replaced{is given by}{has}
\begin{equation}
\eta  = H(Z|U) = \sum _{j =1}^{l}w_{j} h_{m}(T_{ZX}
\boldsymbol{t}_{j}). \label{eq:eta}
\end{equation}
 Let $\mathcal{C}$ be the set of all $(\boldsymbol{p}_{X},s,\eta)$
satisfying (\ref{eq:p}), (\ref{eq:xi}) and (\ref{eq:eta}) for some
choice of $l$, $\boldsymbol{w}$ and $T_{XU}$. Let
$\mathcal{S}$ $=\{(\boldsymbol{p}_{X},h_{n}(T_{YX}
\boldsymbol{p}_{X}),h_{m}(T_{ZX} \boldsymbol{p}_{X})) \in \Delta_{k}
\times [0, \ln n] \times [0,\ln n]\}$.  Each point in $\mathcal{S}$ corresponds to a
$\boldsymbol{p}_{X} \in \Delta_k$.  Thus $\mathcal{C}$ and
$\mathcal{S}$ are both triples whose first term is
$\boldsymbol{p}_{X}$, but the last two terms of $\mathcal{C}$ are
the conditional entropies of $Y$ and $Z$ given $U$ while the last
two terms of $\mathcal{S}$ are the {\em marginal} entropies of $Y$
and $Z$.

Let $\mathcal{C}^*$ $ = \{ (s,\eta ) |
(\boldsymbol{p}_{X},s,\eta) \in $ $\mathcal{C}$ $ \text{ for some }
\boldsymbol{p}_{X}\} $ be the projection of the set $\mathcal{C}$
onto the $(s,\eta)$-plane. Let $\mathcal{C}^*_{\boldsymbol{q}}$ $ =
\{ (s,\eta ) | (\boldsymbol{p}_{X},s,\eta) \in $ $\mathcal{C}$ $,
\boldsymbol{p}_{X} = \boldsymbol{q} \} $ be the subset of $\mathcal{C}^*$  for which  $\boldsymbol{p}_{X}=\boldsymbol{q}$. By definition, $\mathcal{C}^*$ $=\bigcup_{\boldsymbol{q} \in
\Delta_{k}}$ $\mathcal{C}^*_{\boldsymbol{q}}$.

Note that $F^*_{T_{YX},T_{ZX}}(\boldsymbol{q},s)$ is the infimum of
all $\eta$ for which $\mathcal{C}^*_{\boldsymbol{q}}$ contains the
point $(s,\eta)$. Thus
\begin{equation}
F^*_{T_{YX},T_{ZX}}\left(\boldsymbol{q},s\right) = \inf_{\eta}
\left\{ \eta | (\boldsymbol{p}_{X},s,\eta) \in \mbox{$\mathcal{C}$},
\boldsymbol{p}_{X}=\boldsymbol{q}\right\} = \inf_{\eta} \left\{ \eta
| (s,\eta) \in \mathcal{C}^*_{\boldsymbol{q}} \right\}.
\end{equation}

The function $F^*(\boldsymbol{q},s)$ is an extension to DBCs of the function
$F(\boldsymbol{q},s)$ introduced in \cite{Witsenhausen1975}. The
definition of $F(\boldsymbol{q},s)$ is restated here. Let $X
\rightarrow Z$ be a discrete memoryless channel with the $m \times
k$ transition probability matrix $T$, where the entries $T(j,i) =
\text{Pr}(Z=j|X=i)$. Let $\boldsymbol{q}$ be a distribution for
$X$. For any $\boldsymbol{q} \in \Delta _k$, and $0 \leq s \leq
H(X)$, the function $F_{T}(\boldsymbol{q},s)$ is the infimum of
$H(Z|U)$ with respect to all discrete random variables $U$ such that
$H(X|U)=s$ and $U \rightarrow X \rightarrow Z$ is a Markov chain. By
definition, $F_{T}(\boldsymbol{q},s)=F^* _{I,T}(\boldsymbol{q},s)$,
where $I$ is an identity matrix. Most properties of
$F(\boldsymbol{q},s)$ shown in \cite{Witsenhausen1975} can be
\replaced{readily}{straightforwardly} extended to apply to $F^*(\boldsymbol{q},s)$ as well.  These properties
are stated below as propositions. Readers can refer
to \cite{Witsenhausen1975} to see the proofs for $F(\boldsymbol{q},s)$ \replaced{corresponding to}{that easily extend to prove} the propositions for $F^*(\boldsymbol{q},s)$ given below.

\begin{proposition} \label{theorem:CandS}
$\mathcal{C}$ is the convex hull of $\mathcal{S}$. $\mathcal{C}$,
$\mathcal{C}^*$,  and $\mathcal{C}^*_{\boldsymbol{q}}$ are compact,
connected, and convex.  \replaced{See \cite[Section II.A]{Witsenhausen1975}}{See Section II.A of \cite{Witsenhausen1975}}.
\end{proposition}

\begin{proposition} \label{theorem:landk}
i) Every point of $\mathcal{C}$ can be obtained by (\ref{eq:p}),
(\ref{eq:xi}) and (\ref{eq:eta}) with $l \leq k+1$. In other words,
one only need to consider random variables $U$ taking at most $k+1$
values.\\
ii) Every extreme point of the intersection of $\mathcal{C}$ with a
two-dimensional plane can be obtained with $l \leq k$.
\replaced{See \cite[Lemma 2.2]{Witsenhausen1975}.}{See Lemma 2.2 of \cite{Witsenhausen1975}.}
\end{proposition}

\begin{proposition}\label{theorem:srange}
For any fixed $\boldsymbol{q}$ as the distribution of $X$, the
domain of $F^*_{T_{YX},T_{ZX}}(\boldsymbol{q},s)$ in $s$ is the
closed interval $[H(Y|X),H(Y)]=[\sum_{i=1}^{k}q_i
h_{n}(T_{YX}\boldsymbol{e}_i) , h_{n}(T_{YX}\boldsymbol{q})]$, where
$\boldsymbol{e}_i$ is a vector for which the $i^{\textrm{th}}$
entry is 1 and all other entries are zeros.
\end{proposition}

\begin{proof} For the Markov chain $U \rightarrow X \rightarrow Y$,
\replaced{the data processing inequality \cite{BookGallager} implies}{by the Data Processing Theorem \cite{BookGallager},} $H(Y|U) \geq
H(Y|X)$ and equality is achieved when $U=X$. One also has
$H(Y|U) \leq H(Y)$ and equality is achieved when $U$ is a
constant.
\end{proof}

\begin{proposition}\label{theorem:Fdefine}
The function $F^*_{T_{YX},T_{ZX}}(\boldsymbol{q},s)$ is defined and
convex on the compact convex domain
$\{(\boldsymbol{q},s)|\boldsymbol{q} \in \Delta_{k},
\sum_{i=1}^{k}q_i h_{n}(T_{YX}\boldsymbol{e}_i) \leq s \leq
h_{n}(T_{YX}\boldsymbol{q}) \}$ and for each $(\boldsymbol{q},s)$ in
this domain, the infimum in its definition is a minimum, attainable
with $U$ taking at most $k+1$ values.
\replaced{See \cite[Theorem 2.3]{Witsenhausen1975}.}{See Theorem 2.3 of  \cite{Witsenhausen1975}.}
\end{proposition}

\begin{proposition}\label{theorem:Nondecreasing}
$F^*_{T_{YX},T_{ZX}}(\boldsymbol{q},s)$ is monotonically
nondecreasing in $s$ and the infimum in its definition is a minimum.
Hence, $F^*_{T_{YX},T_{ZX}}(\boldsymbol{q},s)$ can be taken as the
minimum $H(Z|U)$ with respect to all \replaced[BX]{$p(u,x)$
satisfying the conditions $H(Y|U)=s$, $\boldsymbol{p}_{X} =
\boldsymbol{q}$, and $U \rightarrow X \rightarrow (Y,Z)$.}{discrete
random variables $U$ such that a) $H(Y|U) \geq s$ and b) $U
\rightarrow X \rightarrow Y,Z$ is a Markov chain.} \replaced{See
\cite[Theorem 2.5]{Witsenhausen1975}.}{See Theorem 2.5 of
\cite{Witsenhausen1975}.}
\end{proposition}

\begin{proposition}\label{theorem:lowerbound}
For any fixed $\boldsymbol{q} =\boldsymbol{p}_{X}$, and $H(Y|X) \leq
s \leq H(Y)$, a lower bound of $F^*(\boldsymbol{q},s)$ is $F^*(\boldsymbol{q},s) \geq s+H(Z)-H(Y)$.
\replaced{See \cite[Theorem 2.6]{Witsenhausen1975}.}{See Theorem 2.6 of \cite{Witsenhausen1975}.}
\end{proposition}

\begin{proposition}\label{theorem:Fregion}
For any given $\boldsymbol{q}=\boldsymbol{p}_{X}$, and  $s$ ranging over the interval  $[H(Y|X),H(Y)]$, the attainable region of $F^*(\boldsymbol{q},s)$ is $H(Z|X) \leq F^*(\boldsymbol{q},s) \leq H(Z) \label{eq:Fregion}$.
\end{proposition}

\begin{proof}
\begin{align} F^*(\boldsymbol{q},s) & = \min_{p(u,x)} \{H(Z|U)|
\boldsymbol{p}_{X}=\boldsymbol{q},H(Y|U)=s \} \nonumber\\
& \geq \min_{p(u,x)} \{H(Z|U,X) |
\boldsymbol{p}_{X}=\boldsymbol{q},H(Y|U)=s \} \label{eq:nondecreasing1}\\
& =H(Z|X),\label{eq:nondecreasing2}
\end{align}
where (\ref{eq:nondecreasing1}) follows \replaced{since conditioning reduces entropy}{from $H(Z|U) \geq H(Z|U,X)$}
and (\ref{eq:nondecreasing2}) follows \replaced{since $Z$ and $U$ are conditionally independent given $X$}{ from the conditional
independence between $Z$ and $U$ given $X$}. Equality is achieved
when $U = X$ and $s = H(Y|X)$. On the other hand,
\begin{align}
F^*(\boldsymbol{q},s) & = \min_{p(u,x)} \{H(Z|U)|
\boldsymbol{p}_{X}=\boldsymbol{q},H(Y|U)=s \} \nonumber\\
& \leq \min_{p(u,x)} \{H(Z) |
\boldsymbol{p}_{X}=\boldsymbol{q},H(Y|U)=s \} \label{eq:nondecreasing3}\\
& =H(Z),
\end{align}
where (\ref{eq:nondecreasing3}) follows \replaced{since conditioning reduces entropy}{from $H(Z|U) \leq H(Z)$}.  Equality is achieved when $U$ is a constant and $s = H(Y)$.
\end{proof}

\begin{proposition}\label{theorem:lambdarange}
For any given $\boldsymbol{q}=\boldsymbol{p}_{X}$, $F^*(s)
\triangleq F^*(\boldsymbol{q},s)$ is differentiable at all but at
most countably many points. At differentiable points of $F^*(s)$,
\begin{equation}
0 \leq \frac{dF^*(s)}{ds} \leq 1 \label{eq:derivative}.
\end{equation}
\end{proposition}

\begin{figure}
  \centering
 \includegraphics[width=0.60\textwidth]{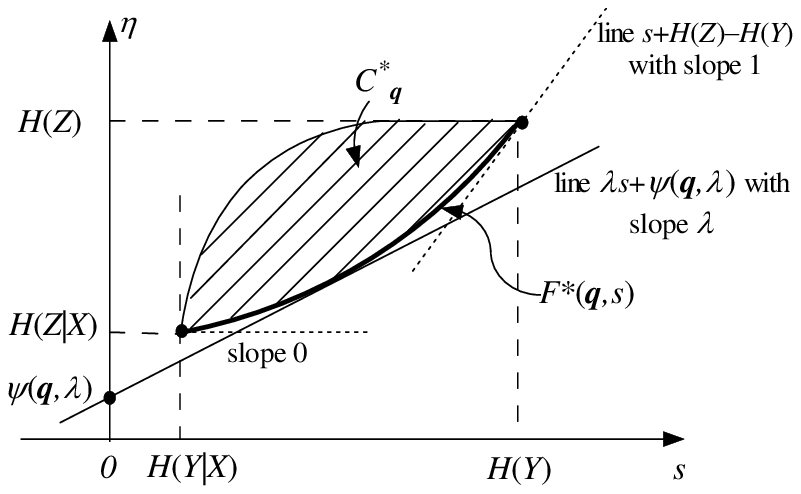}
  \caption{Illustration of the curve $F^*(\boldsymbol{q},s)$ for a given $\boldsymbol{q}$ shown in bold, the region $\mathcal{C}^*_{\boldsymbol{q}}$, and the point $\left(0,\psi(\boldsymbol{q},\lambda)\right)$.}
 \label{fig:Fscratch}
\end{figure}

\begin{proof}
Since $F^*(s)$ is convex in $s$, it is differentiable
at all but at most countably many points. As illustrated in \replaced{Figure}{Fig.}~\ref{fig:Fscratch}, for any $H(Y|X) \leq s
\leq H(Y)$ where $F^*(s)$ is differentiable, the slope of the
supporting line at the point $(s,F^*(s))$ is less than or equal to
the slope of the supporting line $s + H(Z) - H(Y)$ at the point
$(H(Y), F^*(H(Y)))$ because of the convexity of $F^*(s)$. Thus $\frac{dF^*(s)}{ds} \leq 1$ for
any $H(Y|X) \leq s \leq H(Y)$ where $F^*(s)$ is differentiable.  Also, $\frac{dF^*(s)}{ds}
\geq 0$ because $F^*(s)$ is monotonically nondecreasing.
\end{proof}

Let $\boldsymbol{X}=(X_1,\cdots,X_N)$ be a sequence of channel
inputs to the broadcast  channel $X \rightarrow Y
\rightarrow Z$. The
corresponding channel outputs are $\boldsymbol{Y}=(Y_1,\cdots,Y_N)$
and $\boldsymbol{Z}=(Z_1,\cdots,Z_N)$. Thus, any two
channel output pairs $(Y_i,Z_i)$ and $(Y_j,Z_j)$ with $i \ne j$ are
conditionally independent \replaced{given}{with each other given the channel inputs}
$\boldsymbol{X}$. Note that the channel outputs \replaced{$\{(Y_i,Z_i)\}_{i=1}^N$}{$(Y_i,Z_i)$} \replaced{are not necessarily i.i.d.}{do
\emph{not} have to be identically or independently distributed} since
$X_1,\cdots,X_N$ could be correlated and have different
distributions.

Denote
$\boldsymbol{q_i}$ as the distribution of $X_i$ for $i=1,\cdots,N$.
Thus, $\boldsymbol{q}=\sum \boldsymbol{q_i}/N$ is the average of the
distribution of the channel inputs. For any $\boldsymbol{q} \in
\Delta_{k}$, define
$F^*_{T_{YX}^{(N)},T_{ZX}^{(N)}}(\boldsymbol{q},Ns)$ be the infimum
of $H(\boldsymbol{Z}|U)$ with respect to all random variables $U$
and all possible channel inputs $\boldsymbol{X}$ such that
$H(\boldsymbol{Y}|U)=Ns$, the average of the distribution of the
channel inputs is $\boldsymbol{q}$, and $U \rightarrow \boldsymbol{X}
\rightarrow \boldsymbol{Y} \rightarrow \boldsymbol{Z}$ is a Markov
chain.

\begin{proposition}\label{theorem:vectorF}
For all $N=1,2,\cdots,$ and all $T_{YX}$,$T_{ZX}$, $\boldsymbol{q}$,
and $H(Y|X) \leq s \leq H(Y)$, one has $F^*_{T_{YX}^{(N)},T_{ZX}^{(N)}}(\boldsymbol{q},Ns) =
NF^*_{T_{YX},T_{ZX}}(\boldsymbol{q},s). \label{eq:vectorF}$
\replaced{See \cite[Theorem 2.4]{Witsenhausen1975}.}{See Theorem 2.4 of \cite{Witsenhausen1975}.}
\end{proposition}

Proposition \ref{theorem:vectorF} is the key to the applications in Section
\ref{sec:BZC}. \replaced{It indicates that i.i.d. inputs $\boldsymbol{X}$ achieve the conditional entropy bound
$F^*_{T_{YX}^{(N)},T_{ZX}^{(N)}}(\boldsymbol{q},Ns)$.}{It indicates that for a fixed $\boldsymbol{q}$ as
the average of the distribution of the channel inputs, when using the DBC $X \rightarrow Y
\rightarrow Z$ for $N$ times, i.i.d inputs achieve the conditional entropy bound
$F^*_{T_{YX}^{(N)},T_{ZX}^{(N)}}(\boldsymbol{q},Ns)$.}  \added{Moreover,} at each time \added{instant}, a single use of the channel achieves the
conditional entropy bound $F^*_{T_{YX},T_{ZX}}(\boldsymbol{q},s)$.

\begin{theorem}\label{theorem:CapacityRegion}
The capacity region for the discrete memoryless DBC $X \rightarrow Y
\rightarrow Z$ is the closure of the convex hull of all rate pairs
$(R_1,R_2)$ satisfying
\begin{align}
 & 0 \leq R_1 \leq I(X;Y),\label{eq:CapaR1}\\
 & 0 \leq R_2 \leq H(Z) - F^*_{T_{YX},T_{ZX}}\left(\boldsymbol{q},R_1+ H(Y|X)\right),\label{eq:CapaR2}
\end{align}
 for some $\boldsymbol{p}_{X} = \boldsymbol{q} \in \Delta_{k}$,
where $I(X;Y)$, $H(Y|X)$, and $H(Z)$ result from the channel input\deleted{'s}
distribution $\boldsymbol{q}$. \replaced{For a fixed $\boldsymbol{p}_{X}
=\boldsymbol{q}$ and $\lambda \geq 0$, a pareto-optimal rate pair is given by}{Thus, for a fixed $\boldsymbol{p}_{X}
=\boldsymbol{q}$ and for $\lambda \geq 0$, finding the maximum of
$R_2+\lambda R_1$ is equivalent to finding the minimum of
$F^*(\boldsymbol{q},s)-\lambda s$ as follows:}
\begin{align}
\max_{p(u,x)~:~\boldsymbol{p}_{X} = \boldsymbol{q}} \left\{ R_2 + \lambda R_1 \right\} &=H(Z)-\lambda H(Y|X) - \min_{s \in [H(Y|X),H(Y)]} \left\{
F^*\left(\boldsymbol{q},s\right)-\lambda s \right\}. \label{eqn:maxMin}
\end{align}
%
\end{theorem}

\begin{proof} The capacity region for the DBC is known in
\cite{Cover1972} \cite{Gallager1974} \cite{BookCover} as
\begin{equation}
\bar{\text{co}}\left [ \bigcup _{p(u),p(x|u)} \left\{ (R_1,R_2):R_1
\leq I(X;Y|U), R_2 \leq I(U;Z) \right\} \right ],
\label{eq:GallagerCapa}
\end{equation}
where $\bar{\text{co}}$ denotes the closure of the convex hull
operation, and  $U$ is the auxiliary random variable which satisfies
the Markov chain $U \rightarrow X \rightarrow Y \rightarrow Z$ and
$|\mathcal{U}| \leq
\min(|\mathcal{X}|,|\mathcal{Y}|,|\mathcal{Z}|)$. Rewrite
(\ref{eq:GallagerCapa}) and we have
\small
\begin{align}
& \bar{\text{co}}\left [ \bigcup _{p(u),p(x|u)} \left \{
(R_1,R_2):R_1 \leq
I(X;Y|U), R_2 \leq I(U;Z) \right\} \right ] \nonumber \\
= & \bar{\text{co}}\left [ \bigcup
_{\boldsymbol{p}_{X}=\boldsymbol{q} \in \Delta_{k}} \left \{ \bigcup
_{p(u,x) \textrm{~s.t.~} \boldsymbol{p}_{X}=\boldsymbol{q}} \left \{
(R_1,R_2):R_1
\leq I(X;Y|U), R_2 \leq I(U;Z) \right\} \right \} \right ] \label{eq:MyCapa1}\\
= & \bar{\text{co}}\left [ \bigcup
_{\boldsymbol{p}_{X}=\boldsymbol{q} \in \Delta_{k}} \left \{ \bigcup
_{p(u,x) \textrm{~s.t.~} \boldsymbol{p}_{X}=\boldsymbol{q}} \left \{
(R_1,R_2):R_1 \leq H(Y|U) - H(Y|X), R_2 \leq H(Z)- H(Z|U) \right\}
\right \}
\right ] \label{eq:MyCapa2}\\
= & \bar{\text{co}}\left [ \bigcup
_{\boldsymbol{p}_{X}=\boldsymbol{q} \in \Delta_{k}}
\left\{\bigcup_{H(Y|X)\leq s \leq H(Y)}\left \{ (R_1,R_2):R_1 \leq s
- H(Y|X), R_2 \leq H(Z)- F^*_{T_{YX},T_{ZX}}(\boldsymbol{q},s)
\right\}
\right \}\right ] \label{eq:MyCapa3}\\
= & \bar{\text{co}}\left [ \bigcup
_{\boldsymbol{p}_{X}=\boldsymbol{q} \in \Delta_{k}} \left \{
(R_1,R_2): 0 \leq R_1 \leq I(X;Y), R_2 \leq H(Z)-
F^*_{T_{YX},T_{ZX}}(\boldsymbol{q},R_1+ H(Y|X)) \right\} \right ].
\label{eq:MyCapa4}
\end{align}
\normalsize
Some of these steps are justified as follows:
\begin{itemize}
\item (\ref{eq:MyCapa1}) follows from the equivalence of $\bigcup
_{p(u),p(x|u)}$ and $\bigcup _{\boldsymbol{p}_{X}=\boldsymbol{q} \in
\Delta_{k}}
 \bigcup _{p(u,x) \textrm{~s.t.~} \boldsymbol{p}_{X}=\boldsymbol{q}}$;
\item (\ref{eq:MyCapa3}) follows from the definition of the
conditional entropy bound $F^*(\boldsymbol{q},s)$;
\item (\ref{eq:MyCapa4}) follows from the nondecreasing property of
$F^*(s)$ in Proposition \ref{theorem:Nondecreasing}, which allows
the substitution $s=R_1 +H(Y|X)$ in the argument of
$F^*(\boldsymbol{q},s)$.
\end{itemize}
\added{To see that \eqref{eqn:maxMin} holds, observe that:
\begin{align}
& \max_{p(u,x)~:~\boldsymbol{p}_{X} = \boldsymbol{q}} \left\{ R_2 + \lambda R_1 \right\} \nonumber\\
& = \max_{R_1 \in [0,I(X;Y)]}
\left\{ H(Z)-F^*\left(\boldsymbol{q},R_1+H(Y|X)\right) +\lambda
R_1 + \lambda H(Y|X)- \lambda H(Y|X)\right\} \nonumber\\
& =H(Z)-\lambda H(Y|X) + \max_{R_1 \in [0,I(X;Y)]} \left\{ -
F^*\left(\boldsymbol{q},R_1+H(Y|X)\right)+\lambda (R_1+H(Y|X))
\right\} \nonumber \\
& =H(Z)-\lambda H(Y|X) - \min_{s \in [H(Y|X),H(Y)]} \left\{
F^*\left(\boldsymbol{q},s\right)-\lambda s \right\}.\notag
\end{align}}
\end{proof}

Note that for a fixed input distribution $\boldsymbol{q} =
\boldsymbol{p}_{X}$, the items $I(X;Y)$, $H(Z)$ and $H(Y|X)$ in
(\ref{eq:MyCapa4}) are constants. This theorem provides the
relationship between the capacity region and the conditional entropy
bound $F^*(\boldsymbol{q},s)$ for a discrete DBC.

For any given $\boldsymbol{p}_{X} =  \boldsymbol{q}$, \replaced{Theorem
\ref{theorem:CapacityRegion} states that maximizing $R_2+\lambda
R_1$ is equivalent to minimizing
$F^*(\boldsymbol{q},s)-\lambda s$}{by Theorem
\ref{theorem:CapacityRegion}, finding the maximum of $R_2+\lambda
R_1$ is equivalent to finding the minimum of
$F^*(\boldsymbol{q},s)-\lambda s$}. Propositions
\ref{theorem:lowerbound}, \ref{theorem:Fregion}, and
\ref{theorem:lambdarange} indicate that for every $\lambda >1$, the
minimum of $F^*(\boldsymbol{q},s)-\lambda s$ is attained when
$s=H(Y)$ and $F^*(\boldsymbol{q},s)=H(Z)$, i.e., $U$ is a constant.
Thus, the non-trivial range of $\lambda$ is $0 \leq \lambda \leq 1$.

\section{Evaluation of $F^*(\boldsymbol{q},s)$}
\label{sec:evaluation}

In this section, we evaluate $F^*_{T_{YX},T_{ZX}}(\boldsymbol{q},s)$
for a given $\boldsymbol{q}$ via a duality technique, which is also
used for evaluating $F(\cdot)$ in \cite{Witsenhausen1975}. This
duality technique also provides the optimal transmission strategy
for the DBC $X \rightarrow Y \rightarrow Z$ to achieve the maximum
of $R_2 + \lambda R_1$ for any $\lambda \geq 0$.  The section concludes with \replaced{an application to}{the application of the technique to} the binary-symmetric BC.

\subsection{The Duality Technique}

Proposition \ref{theorem:Fdefine} shows that
$F^*_{T_{YX},T_{ZX}}(\boldsymbol{q},s) = \min_{\eta} \{ \eta |
(s,\eta) \in \mbox{$\mathcal{C}$}^{*}_{\boldsymbol{q}} \}$. Thus,
the function $F^*_{T_{YX},T_{ZX}}(\boldsymbol{q},s)$ is determined
by the lower boundary of $\mathcal{C}^*_{\boldsymbol{q}}$ as
illustrated in Figure~\ref{fig:Fscratch}.  Since
$\mathcal{C}^*_{\boldsymbol{q}}$ is convex, its lower boundary can
be described by the lines supporting the boundary from the below.
The line with slope $\lambda$ in the $(s,\eta)$-plane supporting
$\mathcal{C}^*_{\boldsymbol{q}}$ as shown in \replaced{Figure}{Fig.}~\ref{fig:Fscratch}
\replaced{is given by}{has the equation}
\begin{equation}
\eta = \lambda s + \psi (\boldsymbol{q},\lambda),
\label{eq:supporting}
\end{equation}
where $\psi (\boldsymbol{q},\lambda)$ is the $\eta$-intercept of the
tangent line with slope $\lambda$ for the function
$F^*_{T_{YX},T_{ZX}}(\boldsymbol{q},s)$. Thus,
\begin{align}
\psi(\boldsymbol{q},\lambda) & = \min_{s} \left \{
F^*(\boldsymbol{q}, s) - \lambda s \big |
H(Y|X) \leq s \leq H(Y) \right \} \label{eq:intercept1}\\
& = \min_{s,\eta} \left \{ \eta -\lambda s \big | (s, \eta) \in
\mbox{$\mathcal{C}$}^{*}_{\boldsymbol{q}} \right \} \label{eq:intercept2}\\
& = \min_{s,\eta} \left \{ \eta -\lambda s \big | (\boldsymbol{q},
s, \eta) \in \mbox{$\mathcal{C}$} \right \}, \label{eq:intercept3}\\
& = \min_{U \rightarrow X \rightarrow Y,Z \text{ s.t. }
\boldsymbol{p}_{X} = \boldsymbol{q} } \left\{H(Z|U) - \lambda H(Y|U)
\right\}.
\end{align}
For any given $\boldsymbol{q}$, and $H(Y|X) \leq s \leq H(Y)$, the
function $F^*(\boldsymbol{q},s)$ can be represented as
\begin{align}
F^*(\boldsymbol{q},s) & = \max_{\lambda} \{
\psi(\boldsymbol{q},\lambda) +\lambda s
| -\infty < \lambda < \infty \} \label{eq:reconstruct1} \\
& = \max_{\lambda} \{ \psi(\boldsymbol{q},\lambda) +\lambda s | 0
\leq \lambda \leq 1 \}. \label{eq:reconstruct}
\end{align}
where (\ref{eq:reconstruct}) follows from Proposition
\ref{theorem:lambdarange}.

Let $L_{\lambda}$ be \replaced{the}{a} linear transformation
$(\boldsymbol{q},s,\eta) \mapsto (\boldsymbol{q},\eta -\lambda s)$.
\replaced{$L_{\lambda}$}{It} maps $\mathcal{C}$ and $\mathcal{S}$ onto the sets
\begin{equation}
\mbox{$\mathcal{C}$}_{\lambda} = \{ (\boldsymbol{q}, \eta - \lambda
s) | (\boldsymbol{q},s,\eta) \in \mbox{$\mathcal{C}$} \},
\label{eq:Clambda}
\end{equation}
and
\begin{equation}
\mbox{$\mathcal{S}$}_{\lambda} = \{
(\boldsymbol{q},h_{m}(T_{ZX}\boldsymbol{q}) - \lambda
h_{n}(T_{YX}\boldsymbol{q}) )| \boldsymbol{q} \in \Delta_{k} \}.
\label{eq:Slambda}
\end{equation}
Define $\phi(\boldsymbol{q},\lambda) = h_{m}(T_{ZX}\boldsymbol{q}) -
\lambda h_{n}(T_{YX}\boldsymbol{q})$. The lower boundaries of
$\mathcal{C}_{\lambda}$ and $\mathcal{S}_{\lambda}$ are the graphs
of $\psi(\boldsymbol{q},\lambda)$ and $\phi(\boldsymbol{q},\lambda)$
respectively. Since $\mathcal{C}$ is the convex hull of
$\mathcal{S}$, $\mathcal{C}_{\lambda}$ is the convex hull of
$\mathcal{S}_{\lambda}$, and thus $\psi(\boldsymbol{q},\lambda)$ is
the lower convex envelope of $\phi(\boldsymbol{q},\lambda)$ with respect to
$\boldsymbol{q} \in \Delta_{k}$.

\replaced{For each $\lambda$, we conclude that}{In conclusion, for each $\lambda$,} $\psi(\boldsymbol{q}, \lambda)$
can be obtained by forming the lower convex envelope of
$\phi(\boldsymbol{q},\lambda)$ with respect to $\boldsymbol{q}$.
$F^*(\boldsymbol{q},s)$ can be reconstructed from
$\psi(\boldsymbol{q},\lambda)$ by (\ref{eq:reconstruct}). This is
the dual approach to the evaluation of $F^*(\boldsymbol{q},s)$.

Theorem \ref{theorem:CapacityRegion} \replaced{describes}{represents} the capacity region
for a DBC \replaced{in terms of}{by} the function $F^*(\boldsymbol{q},s)$. Since
$\psi(\boldsymbol{q},\lambda)$ and $F^*(\boldsymbol{q},s)$ can be
constructed by each other from (\ref{eq:intercept1}) and
(\ref{eq:reconstruct}) for any $\lambda \geq 0$, the associated
point on the boundary of the capacity region may be found (from its
unique value of $R_2 + \lambda R_1$) as follows
\begin{align}
& \max_{p(u,x)} \{R_2 + \lambda R_1\} \\
= & \max _{\boldsymbol{q} \in \Delta_{k}} \left\{ \max_{p(u,x)
\text{ s.t. } \boldsymbol{p}_{X} = \boldsymbol{q} } \{R_2 + \lambda
R_1 \} \right\}
\nonumber\\
= & \max _{\boldsymbol{q} \in \Delta_{k}} \left\{ \max_{s \in
[H(Y|X),H(Y)],\boldsymbol{p}_{X} = \boldsymbol{q}} \{ H(Z) -
F^*(\boldsymbol{q},s) +\lambda s - \lambda
H(Y|X)\} \right\}\nonumber\\
= & \max _{\boldsymbol{q} \in \Delta_{k}} \left\{ H(Z) - \lambda
H(Y|X) - \min_{s} \{F^*(\boldsymbol{q},s) - \lambda s
\} \big | \boldsymbol{p}_{X} = \boldsymbol{q} \right\} \nonumber\\
= & \max _{\boldsymbol{q} \in \Delta_{k}} \left\{ H(Z) - \lambda
H(Y|X) - \psi(\boldsymbol{q},\lambda) \big | \boldsymbol{p}_{X} =
\boldsymbol{q} \right \}. \label{eq:psiCapa}
\end{align}

We have shown the relationship among $F^*(\boldsymbol{q},s)$,
$\psi(\boldsymbol{q},\lambda)$ and the capacity region for the DBC.
Now we state a theorem which provides the relationship among
$F^*(\boldsymbol{q},s)$, $\psi(\boldsymbol{q},\lambda)$,
$\phi(\boldsymbol{q},\lambda)$, and the optimal transmission
strategies $p(u,x)$ for the DBC. This theorem is a straightforward
extension of Theorem 4.1 in \cite{Witsenhausen1975}.

\begin{theorem}\label{theorem:Fall}
i) For any $0 \leq \lambda \leq 1$, if a point of the graph of
$\psi(\cdot,\lambda)$ is a convex combination of $l$
points of the graph of $\phi(\cdot,\lambda)$ with arguments
$\boldsymbol{t}_{j}$ and
weights $w_{j}$, $j =1,\cdots,l$, then
\begin{equation}
F^*_{T_{YX},T_{ZX}}\left( \sum_{j} w_{j} \boldsymbol{t}_{j},
\sum_{j} w_{j} h_{n}(T_{YX} \boldsymbol{t}_{j})\right) = \sum_{j}
w_{j} h_{m}(T_{ZX} \boldsymbol{t}_{j}). \label{eq:Fall}
\end{equation}
This convex combination representation of a point in
$\psi(\cdot,\lambda)$ implies that for
the fixed channel input distribution
$\boldsymbol{q} = \sum_{j} w_{j}
\boldsymbol{t}_{j}$, an optimal transmission
strategy to achieve the maximum of $R_2 + \lambda R_1$ is determined
by $l$,$w_{j}$ and
$\boldsymbol{t}_{j}$. In
particular, an optimal transmission strategy has $|\mathcal{U}| =$ $
l$, $\text{Pr}(U= j) = w_{j}$ and
$\boldsymbol{p}_{X|U=j} =
\boldsymbol{t}_{j}$,
where
$\boldsymbol{p}_{X|U=j}$ denotes the conditional distribution of $X$ given $U = j$. \\
ii)For a predetermined channel input distribution $\boldsymbol{q}$,
if the transmission strategy $|\mathcal{U}| =$ $ l$, $\text{Pr}(U=
j) = w_{j}$ and $\boldsymbol{p}_{X|U=j} =
\boldsymbol{t}_{j}$
achieves $\max \{ R_2 +\lambda R_1| \sum_{j} w_{j}
\boldsymbol{t}_{j} = \boldsymbol{q} \}$, then the
point $(\boldsymbol{q},\psi(\boldsymbol{q},\lambda))$ is the convex
combination of $l$ points of the graph of $\phi(\cdot,\lambda)$ with
arguments $\boldsymbol{t}_{j}$
and weights $w_{\lambda}$, $j =1,\cdots,l$.
\end{theorem}

Note that if for some pair $(\boldsymbol{q},\lambda)$,
$\psi(\boldsymbol{q},\lambda) = \phi(\boldsymbol{q},\lambda)$, then
the corresponding optimal transmission strategy has $l=1$, which
means $U$ is a constant. For such a $(\boldsymbol{q},\lambda)$ pair, the
line $\eta = \lambda s + \psi(\boldsymbol{q},\lambda)$ supports the
graph of $F^*(s)$ at its endpoint
$(H(Y),H(Z))=(h_n(T_{YX}\boldsymbol{q}),h_m(T_{ZX}\boldsymbol{q}))$.

\subsection{Example: Application to the binary-symmetric broadcast channel} \label{sec:BSC}
Consider the binary-symmetric BC $X \rightarrow Y \rightarrow Z$ with
\begin{equation} \label{eq:BBSC}
T_{YX}=\begin{bmatrix}
1- \alpha_1 & \alpha_1\\
\alpha_1 & 1- \alpha_1
\end{bmatrix},
T_{ZX}=\begin{bmatrix}
1- \alpha_2 & \alpha_2\\
\alpha_2 & 1- \alpha_2
\end{bmatrix},
\end{equation}
where $0 < \alpha_1 < \alpha_2 < 1/2$.  The following theorem, which is proved by the duality technique, provides an explicit parametrized characterization of the capacity region.

\begin{theorem} \label{theorem:BSCcapRegion}
Consider the binary symmetric BC with crossover probabilities $0 < \alpha_1 < \alpha_2 < 1/2$. For $\lambda\geq 0$, the achievable rate pair $(R_1,R_2)$ which maximizes $\lambda R_1+R_2$ is given by
\begin{align*}
 R_1 &= h\left ( \alpha_1 + (1-2\alpha_1)p_{\lambda}\right) -h(\alpha_1), \\
 R_2 &= \ln(2) - h\left ( \alpha_2 +(1-2\alpha_2)p_{\lambda}\right),
\end{align*}
where $\lambda$, $R_1$, and $R_2$ are parametrized by $0\leq p_{\lambda}\leq 1/2$ satisfying
\begin{align*}
\lambda = \frac{1-2\alpha_2}{1-2\alpha_1} \cdot
\frac{\ln{\frac{1-\alpha_2-(1-2\alpha_2)p_{\lambda}}{\alpha_2+(1-2\alpha_2)p_{\lambda}}}}
{\ln{\frac{1-\alpha_1-(1-2\alpha_1)p_{\lambda}}{\alpha_1+(1-2\alpha_1)p_{\lambda}}}}.
\end{align*}
Moreover, NE achieves all points in the capacity region.
\end{theorem}

\begin{figure}
  \centering
  \includegraphics[width=0.80\textwidth]{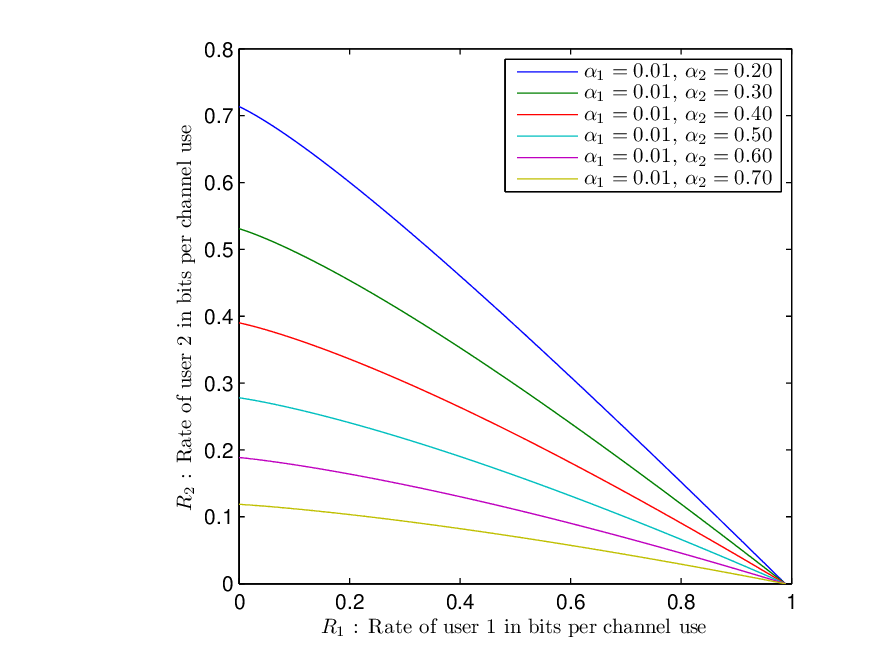}
  \caption{Binary symmetric broadcast channel capacity regions (in bits per channel use) obtained using the explicit parametric expressions given in Theorem \ref{theorem:BSCcapRegion} for $\alpha_1=0.001$ and  a variety of $\alpha_2$ values. }\label{fig:DBCplot}
\end{figure}

Figure \ref{fig:DBCplot} shows several example capacity region boundaries computed using Theorem \ref{theorem:BSCcapRegion}.

\begin{proof}
For the binary-symmetric BC $X \rightarrow Y \rightarrow Z$ with $0 < \alpha_1 < \alpha_2 < 1/2$, one has
\begin{align}
\phi(p,\lambda) & \overset{\Delta}{=} \phi\left([p,1-p]^T,\lambda
\right)
\nonumber\\
& = h_{m}\left(T_{ZX} \boldsymbol{q}\right) - \lambda
h_{n}\left(T_{YX}
\boldsymbol{q} \right) \nonumber\\
& = h\left( ( 1 - \alpha_2 ) p + \alpha_2 ( 1 - p )\right) - \lambda
h\left( (1-\alpha_1 ) p + \alpha_1 (1-p)\right).
\end{align}
Taking the second derivative of $\phi(p,\lambda)$ with respect to
$p$, we have
\begin{align}
\phi '' (p,\lambda) = \frac{-(1-2 \alpha_2)^2}{\left(\alpha_2 p
+(1-\alpha_2)(1-p)\right)\left((1-\alpha_2)p + \alpha_2 (1-p)\right)} \nonumber\\
+ \frac{ \lambda (1-2 \alpha_1)^2}{\left(\alpha_1 p
+(1-\alpha_1)(1-p)\right) \left((1-\alpha_1)p + \alpha_1
(1-p)\right)}\, . \label{eq:phi''}
\end{align}
In \eqref{eq:phi''}, $\phi '' (p,\lambda) =-A + \lambda B$ where $A$ and $B$ are both positive.  Thus  $\phi '' (p,\lambda)$ has the sign of
\begin{equation}
\rho (p,\lambda) = \frac{\phi '' (p,\lambda)}{AB}= -\left(\frac{1-\alpha_1}{1-2\alpha_1} -
p \right) \left(\frac{\alpha_1}{1-2\alpha_1} + p \right) + \lambda
\left(\frac{1-\alpha_2}{1-2\alpha_2} - p \right) \left(\frac{\alpha_2}{1-2\alpha_2}
+ p \right).
\end{equation}
For any $0 \leq \lambda \leq 1$, $p=1/2$ minimizes $\rho$ so that
\begin{equation}
\min_{p} \rho (p,\lambda) = \frac{\lambda}{4(1-2\alpha_2)^2} -
\frac{1}{4(1-2\alpha_1)^2}.
\end{equation}
Thus, for $\lambda \geq (1-2\alpha_2)^2/(1-2\alpha_1)^2$, $\phi
''(p,\lambda) \geq 0$ for all $0 \leq p \leq 1$, and so
$\psi(p,\lambda) = \phi(p,\lambda)$. In this case, the transmission strategy that maximizes $R_1$ also maximizes $R_2 + \lambda R_1$.  Thus, the optimal
transmission strategy has $l=1$, which means $U$ is a constant.

Note that $\phi(1/2+p,\lambda) = \phi(1/2 -p,\lambda)$. For $\lambda
< (1-2\alpha_2)^2/(1-2\alpha_1)^2$, $\phi(p,\lambda)$ has negative
second derivative on an interval symmetric about $p=1/2$. Let
$p_{\lambda} = \arg \min_{p} \phi(p,\lambda)$ with $p_{\lambda} \leq
1/2$. Thus $p_{\lambda}$ satisfies $\phi_{p} ' (p_{\lambda},\lambda)
= 0$.

\begin{figure}
  \centering
  \includegraphics[width=0.4\textwidth]{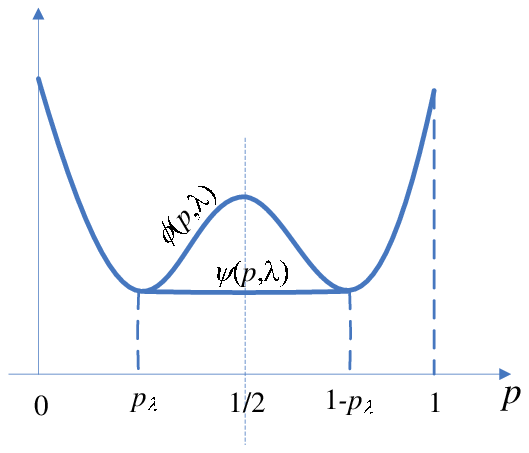}
  \caption{An illustration of $\psi(p,\lambda)$ and $\phi(p,\lambda)$ for the binary symmetric BC with $\lambda
< (1-2\alpha_2)^2/(1-2\alpha_1)^2$. }\label{fig:psi_binarysym}
\end{figure}

By symmetry, the envelope $\psi(\cdot, \lambda)$ is obtained by
replacing $\phi(p,\lambda)$ on the interval $(p_{\lambda},
1-p_{\lambda})$ by its minimum over $p$, as shown in
Figure~\ref{fig:psi_binarysym}. Therefore, the lower envelope of
$\phi(p,\lambda)$  for the binary symmetric BC is
\begin{equation}
\psi(p,\lambda) = \left \{
\begin{array}{ll}
\phi(p_{\lambda},\lambda), & \textrm{for } p_{\lambda} \leq p \leq
1-p_{\lambda} \\
 \phi(p,\lambda), & \textrm{otherwise}.
\end{array} \right .
\end{equation}

For  a predetermined distribution of $X$,
$\boldsymbol{p}_{X}=\boldsymbol{q}=[q,1-q]^{T}$ with $p_{\lambda} <
q < 1-p_{\lambda}$, the pair $(q, \psi(q, \lambda) )$ is the convex
combination of the points $(p_{\lambda}, \phi(p_{\lambda},\lambda))$
and $(1-p_{\lambda},\phi(1-p_{\lambda},\lambda))$. Therefore, by
Theorem \ref{theorem:Fall}, the optimal transmission strategy
\deleted{constraint} with $\boldsymbol{p}_{X}=\boldsymbol{q}$ is \added{NE with}
\begin{equation}
\boldsymbol{p}_{U}=\left [
                      \begin{array}{c}
                        \frac{1-p_{\lambda}-q}{1-2 p_{\lambda}}\\
                        \frac{q-p_{\lambda}}{1-2 p_{\lambda}}\\
                      \end{array}
                    \right ]
\text{ and } T_{XU}=\left [
                      \begin{array}{cc}
                        p_{\lambda} & 1-p_{\lambda}\\
                        1-p_{\lambda} & p_{\lambda}\\
                      \end{array}
                    \right ]. \label{eq:BBSCopt}
\end{equation}
The conditional entropy bound
$F^*(\boldsymbol{q},s)=h_{2}(T_{ZX}\cdot
[p_{\lambda},1-p_{\lambda}]^{T})=h(\alpha_2+(1-2\alpha_2)p_{\lambda})$
for $s=h_{2}(T_{YX}\cdot
[p_{\lambda},1-p_{\lambda}]^{T})=h(\alpha_1+(1-2\alpha_1)p_{\lambda})$,
and $p_{\lambda} \leq q \leq 1-p_{\lambda}$. For the given
$\boldsymbol{q}$, this defines $F^*(s) \triangleq
F^*(\boldsymbol{q}, s)$ on its entire domain $s \in [h(\alpha_1),
h(\alpha_1+(1-2\alpha_1)q)]$, i.e., $s \in$ $[H(Y|X),
H(Y)]$.

Note that for a predetermined distribution of $X$,
$\boldsymbol{p}_{X}=\boldsymbol{q}=[q,1-q]^{T}$ with the suboptimal choices of $q < p_{\lambda}$ or $q
>1-p_{\lambda}$, one has $\phi(q,\lambda)=\psi(q,\lambda)$, which means that
a line with slope $\lambda$ supports $F^*(\boldsymbol{q},\cdot)$ at
point $s = H(Y)= h(\alpha_1+(1-2\alpha_1)q)$, and thus the optimal
transmission strategy \replaced{under the constraint that $q < p_{\lambda}$ or $q
>1-p_{\lambda}$}{constraint with
$\boldsymbol{p}_{X}=\boldsymbol{q}$} has $l=1$, which means $U$ is a
constant.

The boundary of the capacity region for the binary-symmetric BC is
always achieved when $\boldsymbol{p}_{X} = [1/2,1/2]^{T}$ (see
\cite{Bergmans1973}). Hence, the optimal transmission strategy to
achieve the boundary of the capacity region always has $l=2$ and
follows from (\ref{eq:BBSCopt}) with $q=1/2$.  This leads to the
following explicit parametric expression for the boundary of the
capacity region of the two-receiver binary-symmetric BC:
\begin{align}
 & R_1 = h\left ( \alpha_1 + (1-2\alpha_1)p_{\lambda}\right) -h(\alpha_1), \label{eq:BSBCcapa1}\\
 & R_2 = \ln(2)- h\left ( \alpha_2 +(1-2\alpha_2)p_{\lambda}\right),\label{eq:BSBCcapa2}
\end{align}
where the parameter $p_{\lambda}$ is ranging from 0 to 1/2. In
addition, the rate pair $(R_1, R_2)$ in (\ref{eq:BSBCcapa1}) and
(\ref{eq:BSBCcapa2}) maximizes $R_2+\lambda R_1$ for each pair of
$\lambda$ and $p_{\lambda}$ satisfying $\phi_{p} '
(p_{\lambda},\lambda) = 0$, which implies
\begin{align*}
\lambda = \frac{1-2\alpha_2}{1-2\alpha_1} \cdot
\frac{\ln{\frac{1-\alpha_2-(1-2\alpha_2)p_{\lambda}}{\alpha_2+(1-2\alpha_2)p_{\lambda}}}}
{\ln{\frac{1-\alpha_1-(1-2\alpha_1)p_{\lambda}}{\alpha_1+(1-2\alpha_1)p_{\lambda}}}}.
\end{align*}
\end{proof}

\section{Broadcast Z Channels}
\label{sec:BZC}

\begin{figure}
  \centering
  \includegraphics[width=0.60\textwidth]{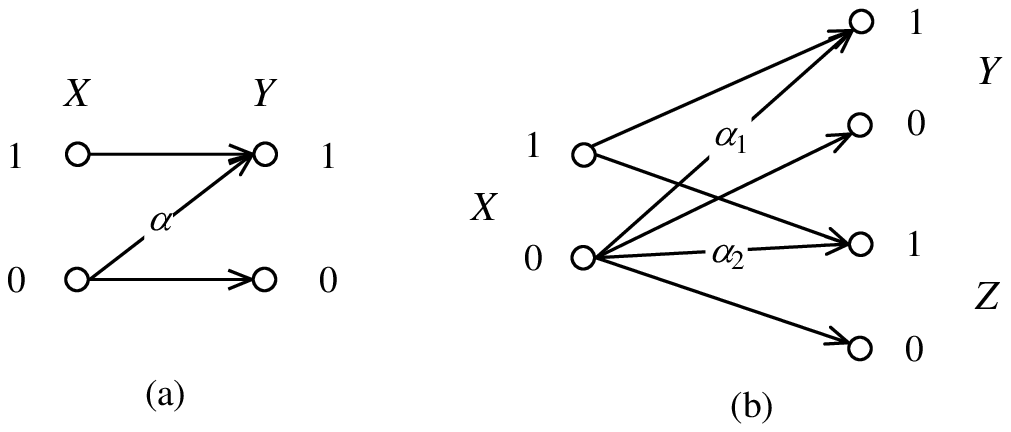}
  \caption{The Z channel (a) and broadcast Z channel (b).}\label{fig:BZChannel}
\end{figure}

\begin{figure}
  \centering
  \includegraphics[width=0.30\textwidth]{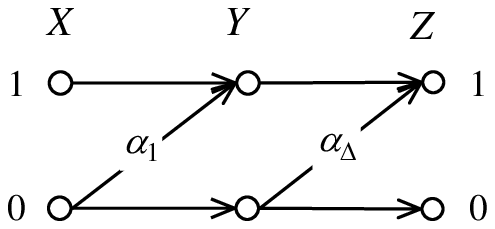}
  \caption{A physically degraded broadcast Z channel.}\label{fig:DBZC}
\end{figure}

The Z channel, shown in Figure~\ref{fig:BZChannel}(a), is a binary
asymmetric channel which is noiseless when symbol 1 is transmitted
but noisy when symbol 0 is transmitted. The channel output $Y$ is
the binary OR of the channel input $X$ and Bernoulli distributed
noise with parameter $\alpha$. The capacity of the Z channel was
studied in \cite{Golomb1980}. The Broadcast Z channel is a class of
discrete memoryless broadcast channels whose component channels are
Z channels. A two-receiver broadcast Z channel with marginal
transition probability matrices
\begin{equation}
T_{YX}=\begin{bmatrix}
1 & \alpha_1\\
0 & 1- \alpha_1
\end{bmatrix},
T_{ZX}=\begin{bmatrix}
1 & \alpha_2\\
0 & 1- \alpha_2
\end{bmatrix},
\end{equation}
where $0 < \alpha_1 \leq \alpha_2 < 1$, is shown in
Fig~\ref{fig:BZChannel}(b). The two-receiver broadcast Z channel is
stochastically degraded and can be modeled as a physically degraded
broadcast channel as shown in Figure~\ref{fig:DBZC}, where
$\alpha_{\Delta}= (\alpha_2 -\alpha_1)/(1-\alpha_1)$ \cite{BZC}.
NE for broadcast Z channels uses the binary OR function to
combine each receiver's independently encoded message.
As shown in \cite{BZC} \cite{Bike2007}, NE achieves the entire boundary of the
capacity region for the two-receiver broadcast Z channel.  In this
section, we will show that  NE also
achieves the entire boundary of the capacity region for
 broadcast Z channels with more than
two receivers.

\subsection{\replaced{Capacity region}{$F^*$} for the \added{two-receiver} broadcast Z channel}
Similar to Theorem \ref{theorem:BSCcapRegion} for the BS broadcast channel, we can apply our analysis of $F^*$ to obtain a parametric expression for the capacity region of the broadcast Z channel.
\begin{theorem} \label{theorem:BZcapRegion}
Consider the broadcast Z channel with crossover probabilities $0 <
\alpha_1 \leq \alpha_2 < 1$. Define $\beta_i = 1- \alpha_i$ for
$i=1,2$. For $\lambda\geq 0$, the achievable rate pair $(R_1,R_2)$
which maximizes $\lambda R_1+R_2$ is given by \added[BX]{
\begin{align}
R_1 &= \frac{q_{\lambda}}{p_{\lambda}}h(\beta_1 p_{\lambda}) -
q_{\lambda}
h(\beta_1),\label{eqn:BZCcapa1_q}\\
R_2 &= h(q_{\lambda} \beta_2) -
\frac{q_{\lambda}}{p_{\lambda}}h(\beta_2
p_{\lambda}),\label{eqn:BZCcapa2_q}
\end{align}}
where $\lambda$,\added[BX]{ $q_{\lambda}$,} $R_1$, and $R_2$ are
parametrized by \replaced[BX]{$0\leq p_{\lambda}\leq 1$}{$0\leq
q\leq p_{\lambda}\leq 1$} satisfying
\begin{align}
\lambda &= \frac{\ln (1- \beta_2 p_{\lambda})}{\ln(1-\beta_1
p_{\lambda})} \label{eqn:BZCplambda}\\
q_{\lambda} &= \min \left(p_{\lambda},\frac{1}{\beta_2
\left(1+\exp\left({\frac{1}{\beta_2 p_{\lambda}} \left(h(\beta_2
p_{\lambda}) - \lambda h(\beta_1 p_{\lambda}) + \lambda p_{\lambda}
h(\beta_1) \right)}\right) \right)} \right).\label{eqn:BZCqlambda}
\end{align}
Moreover, NE achieves all points in the capacity region.
\end{theorem}

Thus, Theorem \ref{theorem:BZcapRegion} implies that for a specified $\alpha_1$ and $\alpha_2$, the capacity region for the two-receiver broadcast Z channel can be determined parametrically for each $\lambda$ as follows:
\begin{enumerate}
\item Use (\ref{eqn:BZCplambda}) to compute $p_{\lambda}$ from $\lambda$.
\item Use (\ref{eqn:BZCqlambda}) to compute $q_{\lambda}$ from $p_{\lambda}$.
\item Use $q_{\lambda}$ and $p_{\lambda}$ in (\ref{eqn:BZCcapa1_q}) and (\ref{eqn:BZCcapa2_q}) to find the $R_1$ and $R_2$ that maximize $R_2+\lambda R_1$.
\end{enumerate}

\begin{figure}
  \centering
  \includegraphics[width=0.80\textwidth]{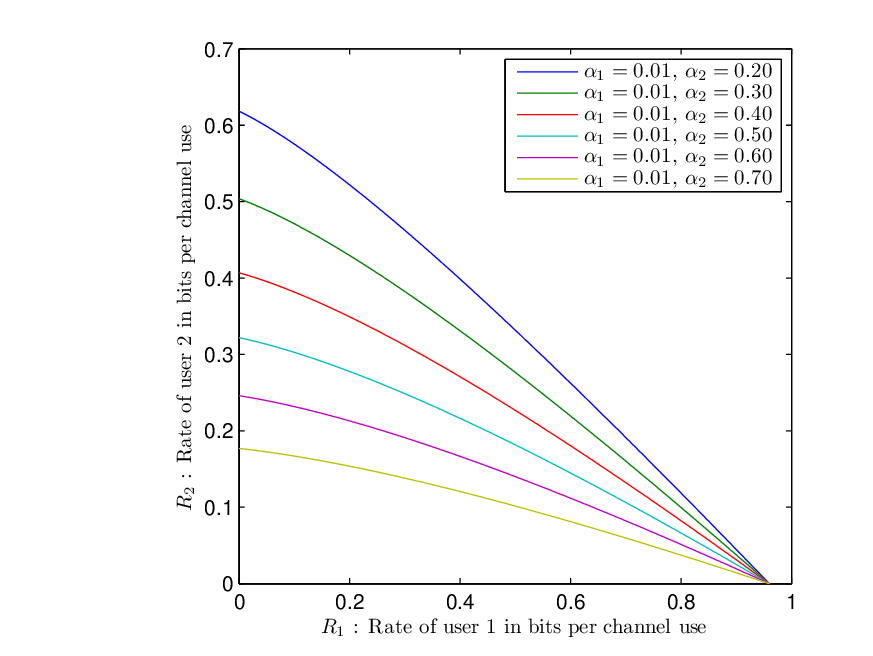}
  \caption{Broacast Z channel channel capacity regions (in bits per channel use) obtained using the explicit parametric procedure for $\alpha_1=0.01$ and  a variety of $\alpha_2$ values. }\label{fig:BZCplot}
\end{figure}

Figure \ref{fig:BZCplot} shows several example capacity region boundaries found using this procedure.

\begin{figure}
  \centering
  \includegraphics[width=0.45\textwidth]{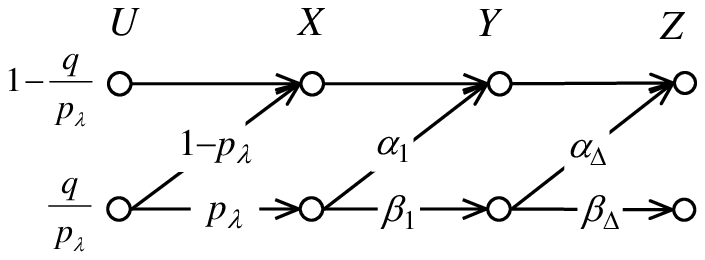}
  \caption{An optimal transmission strategy for the two-receiver broadcast Z channel.}\label{fig:opt_Z}
\end{figure}

\begin{proof}
For the broadcast Z channel $X \rightarrow Y \rightarrow Z$ shown in
Figure~\ref{fig:BZChannel}(b) and Figure~\ref{fig:DBZC} with
\begin{equation} \label{eq:BZC}
T_{YX}=\begin{bmatrix}
1 & \alpha_1\\
0 & \beta_1
\end{bmatrix},
T_{ZX}=\begin{bmatrix}
1 & \alpha_2\\
0 & \beta_2
\end{bmatrix},
\end{equation}
where $0 < \alpha_1 \leq \alpha_2 < 1$, $\beta_1 = 1- \alpha_1$, and
$\beta_2= 1- \alpha_2$, one has
\begin{equation}
\phi(p,\lambda) \overset{\Delta}{=} \phi \left([1-p,p]^T,\lambda
\right)=h(p\beta_2)-\lambda h(p \beta_1). \label{eq:phi_BZC}
\end{equation}
Taking the second derivative of $\phi(p, \lambda)$ with respect to
$p$, we have

\begin{equation}
\phi ''(p,\lambda)= \frac{-\beta_2}{(1-p\beta_2)p} +
\frac{\lambda \beta_1}{(1-p\beta_1)p}, \label{eq:BZCphi''}
\end{equation}

Multiplying $\phi '' (p,\lambda)$ in \eqref{eq:BZCphi''}
by the positive quantity $(1-p\beta_1)(1-p\beta_2)p$ produces
\begin{equation}
\rho (p,\lambda) =\phi '' (p,\lambda) \cdot
(1-p\beta_1)(1-p\beta_2)p = p\beta_1 \beta_2 (1-\lambda) + \lambda
\beta_1 - \beta_2,
\end{equation}
which has the same sign as $\phi '' (p,\lambda)$.

Let $\beta_{\Delta} \overset{\Delta}{=}\beta_2/\beta_1$. For the
case of $\beta_{\Delta}  \leq \lambda \leq 1$, $\phi ''(p,\lambda)
\geq 0$ for all $0 \leq p \leq 1$. Hence, $\phi(p,\lambda)$ is
convex in $p$  and thus $\phi(p,\lambda) = \psi(p,\lambda)$ for all
$0 \leq p \leq 1$. In this case, the transmission strategy that maximizes $R_1$ also maximizes $R_2 + \lambda R_1$.  Thus, the optimal transmission strategy has $l=1$, i.e.,
$U$ is a constant.  Note that the transmission strategy with $l=1$ is
a special case of the NE scheme in which the only codeword for the
second receiver is an all-ones codeword.

\begin{figure}
  \centering
  \includegraphics[width=0.4\textwidth]{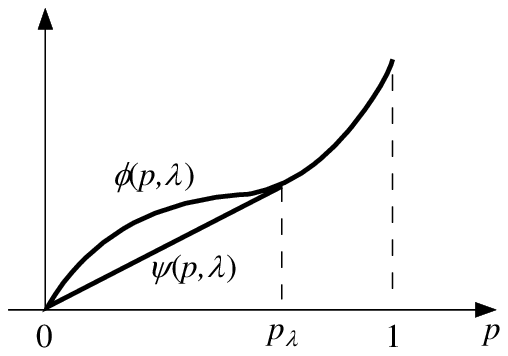}
  \caption{Illustration of $\phi(p,\lambda)$ and $\psi(p,\lambda)$ for the broadcast Z channel with a given $\lambda$.}\label{fig:psi_Z}
\end{figure}

For the case of $0 \leq \lambda < \beta_{\Delta}$, $\phi
(p,\lambda)$ is concave in $p$ on $[0, \frac{\beta_2 - \lambda
\beta_1}{\beta_1 \beta_2 (1- \lambda)}]$ and convex on
$[\frac{\beta_2 - \lambda \beta_1}{\beta_1 \beta_2 (1-
\lambda)},1]$. Figure~\ref{fig:psi_Z} illustrates the graph
in this case. Since $\phi(0,\lambda) = 0$,
$\psi(\cdot,\lambda)$, the lower convex envelope of
$\phi(\cdot,\lambda)$, is constructed using the tangent of
$\phi(\cdot,\lambda)$ that passes  through
the origin as shown in Figure~\ref{fig:psi_Z}. Let
$(p_{\lambda}, \phi(p_{\lambda},\lambda))$ be the point of contact.
The value of $p_{\lambda}$ is determined by
$\phi_{p}'(p_{\lambda},\lambda) =
\phi(p_{\lambda},\lambda)/p_{\lambda}$, i.e.,
\begin{equation}
\lambda = \frac{\ln (1- \beta_2 p_{\lambda})}{\ln(1-\beta_1
p_{\lambda})}. \label{eq:BZCplambda}
\end{equation}

Let $\boldsymbol{q} = [1-q,q]^T$ be the distribution of the channel
input $X$. For $q \leq p_{\lambda}$, $\psi(q,\lambda)$ is obtained
as a convex combination of points $(0,0)$ and
$(p_{\lambda},\phi(p_{\lambda},\lambda))$ with weights
$(p_{\lambda}-q)/p_{\lambda}$ and $q/p_{\lambda}$. By Theorem
\ref{theorem:Fall}, it corresponds to
$s=[(p_{\lambda}-q)/p_{\lambda}]\cdot 0+[q/p_{\lambda}] \cdot
h(\beta_1 p_{\lambda}) = q h(\beta_1 p_{\lambda})/p_{\lambda}$ and
$F^*(q,s)\triangleq F^*(\boldsymbol{q},s) = q/p_{\lambda} \cdot
h(\beta_2 p_{\lambda})$. Hence, for the broadcast Z channel,
\begin{equation}
F^*_{T_{YX},T_{ZX}}(q,q h(\beta_1 p)/p)= q h(\beta_2 p)/p
\label{eq:F_Z}
\end{equation}
for $p \in [q,1]$, which defines $F^*_{T_{YX},T_{ZX}}(q,\cdot)$ on
its entire domain $[q h(\beta_1), h(q \beta_1)]$. Also by Theorem
\ref{theorem:Fall}, the optimal transmission strategy $p(u,x)$
to maximize $(R_2+\lambda R_1)$ given the
constraint $\boldsymbol{p}_{X} = \boldsymbol{q}$
 is
determined by $l=2$, $w_1=(p_{\lambda}-q)/p_{\lambda}$, $w_2 =
q/p_{\lambda}$, $\boldsymbol{t}_1=[1,0]^T$ and
$\boldsymbol{t}_2=[1-p_{\lambda},p_{\lambda}]^T$. Since the
optimal transmission strategy $p(u,x)$ can be modeled as a Z
channel as shown in Figure~\ref{fig:opt_Z}, the random variable $X$
can be constructed as the OR
 of two Bernoulli random variables with
parameters $(p_{\lambda}-q)/p_{\lambda}$ and $1-p_{\lambda}$
respectively. Hence, an optimal transmission
strategy for the broadcast Z channel is NE. For $q > p_{\lambda}$, $\psi(q,\lambda) =
\phi(q, \lambda)$ and  an optimal strategy has $l=1$, i.e., $U$ is a constant.  \deleted{Thus
the $F^*$ approach provides an alternative proof that NE achieves
the entire capacity region boundary for the two-receiver broadcast Z channel.}

Thus, the two-receiver broadcast Z channel capacity
region is the convex hull of the rate pairs $(R_1,R_2)$
satisfying
\begin{align}
& 0 \leq R_1 \leq \frac{q}{p_{\lambda}}h(\beta_1 p_{\lambda}) - q
h(\beta_1), \\
& 0 \leq R_2 \leq h(q \beta_2) - \frac{q}{p_{\lambda}}h(\beta_2
p_{\lambda}),
\end{align}
for some $q \in [0,1]$ and  $p_{\lambda} \in [q,1]$.  For a fixed input distribution $\boldsymbol{p}_{X} = [1-q,q]^{T}$, the rate
pair $(R_1, R_2)$ of
\begin{align}
& R_1 = \frac{q}{p_{\lambda}}h(\beta_1 p_{\lambda}) - q
h(\beta_1), \label{eq:BZCcapa1_q}\\
& R_2 = h(q \beta_2) - \frac{q}{p_{\lambda}}h(\beta_2
p_{\lambda}),\label{eq:BZCcapa2_q}
\end{align}
maximizes $R_2+\lambda R_1$ for each pair of $\lambda$ and
$p_{\lambda}$ satisfying (\ref{eq:BZCplambda}). Among all possible
input distribution\added{s} $q \in [0,1]$, only one will finally maximize
$R_2+\lambda R_1$ over all rate pairs in the capacity region. Let
$q_{\lambda}$ be the input distribution which maximizes $R_2+\lambda
R_1$, and thus,
\begin{align}
q_{\lambda} & = \arg \max_{0 \leq q \leq p_{\lambda}}(R_2+ \lambda R_1) \\
            & = \arg \max_{0 \leq q \leq p_{\lambda}}\left( h(q \beta_2) - \frac{q}{p_{\lambda}}h(\beta_2
p_{\lambda}) + \lambda \left( \frac{q}{p_{\lambda}}h(\beta_1
p_{\lambda}) - q h(\beta_1) \right) \right),\\
& = \min \left(p_{\lambda},\frac{1}{\beta_2
\left(1+\exp\left({\frac{1}{\beta_2 p_{\lambda}} \left(h(\beta_2
p_{\lambda}) - \lambda h(\beta_1 p_{\lambda}) + \lambda p_{\lambda}
h(\beta_1) \right)}\right) \right)} \right). \label{eq:BZCqlambda}
\end{align}
\end{proof}

\subsection{The broadcast Z channel with more than two receivers}


\begin{figure}
  \centering
  \includegraphics[width=0.60\textwidth]{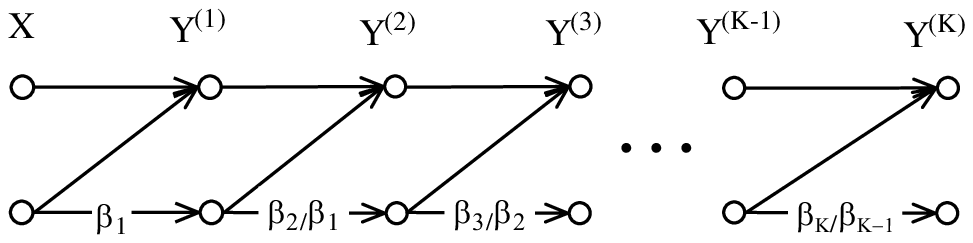}
  \caption{The $K$-receiver broadcast Z channel}\label{fig:DKBZC}
\end{figure}

Consider a $K$-receiver broadcast Z channel $X \rightarrow Y^{(1)}
\rightarrow \cdots \rightarrow Y^{(K)}$ with marginal transition
probability matrices
\begin{equation}
T_{Y_{j}X}=\begin{bmatrix}
1 & \alpha_j\\
0 & \beta_j
\end{bmatrix},
\end{equation}
where $0 < \alpha_1 \leq \cdots \leq \alpha_K < 1$, and $\beta_j =
1- \alpha_j$ for $j=1,\cdots,K$. The $K$-receiver broadcast Z
channel is stochastically degraded and can be modeled as a
physically DBC as shown in Figure~\ref{fig:DKBZC}. NE for the $K$-receiver broadcast Z channel combines the $K$ independently generated codewords (one for each receiver) using the binary OR operation.  The
$j^{\textrm{th}}$ receiver then successively decodes the messages
for Receiver $K$, Receiver $K-1$, $\cdots$, and finally for Receiver
$j$. The codebook for the $j^{\textrm{th}}$ receiver is \replaced{a random codebook drawn}{designed by
random coding technique} according to the binary random variable
$X^{(j)}$ with $\text{Pr}\{ X^{(j)} = 0\} = q^{(j)}$. Denote
$X^{(i)} \circ X^{(j)}$ as the binary OR of $X^{(i)}$ and $X^{(j)}$.
Hence, the channel input $X$ is the OR of $X^{(j)}$ for all $1 \leq
j \leq K$, i.e., $X= X^{(1)} \circ \cdots \circ X^{(K)}$. From the analysis of successive decoding in the proof of the
coding theorem for DBCs \cite{Bergmans1973} \cite{Gallager1974}, the
achievable region of NE for the $K$-receiver broadcast Z
channel is determined by
\begin{align}
R_{j} & \leq I \left( Y^{(j)},X^{(j)}|X^{(j+1)},\cdots, X^{(K)} \right) \\
 & = H \left( Y^{(j)}|X^{(j+1)},\cdots, X^{(K)} \right) -  H \left( Y^{(j)}|X^{(j)}, X^{(j+1)},\cdots, X^{(K)} \right) \\
 & = \left (\prod_{i=j+1}^{K}q^{(i)} \right ) \cdot h \left( \beta_{j} \prod_{i=1}^{j}q^{(i)} \right)  -
\left (\prod_{i=j}^{K}q^{(i)}\right ) \cdot h \left( \beta_{j} \prod_{i=1}^{j-1}q^{(i)}  \right) \\
& = \frac{q}{t_j}h(\beta_{j}t_j) -
\frac{q}{t_{j-1}}h(\beta_{j}t_{j-1}), \label{eq:AchReg_KBZC}
\end{align}
where $t_j = \prod_{i=1}^{j}q^{(i)} $ for $j=1,\cdots,K$, and $q=
\textrm{Pr}(X=0)= \prod_{i=1}^{K}q^{(i)}$. Denote $t_0=1$. Since $0
\leq q^{(1)},\cdots, q^{(K)} \leq 1$, one has
\begin{equation}
1 = t_0 \geq t_1 \geq \cdots \geq t_K = q. \label{eq:t_define}
\end{equation}

\begin{figure}
  \centering
  \includegraphics[width=0.80\textwidth]{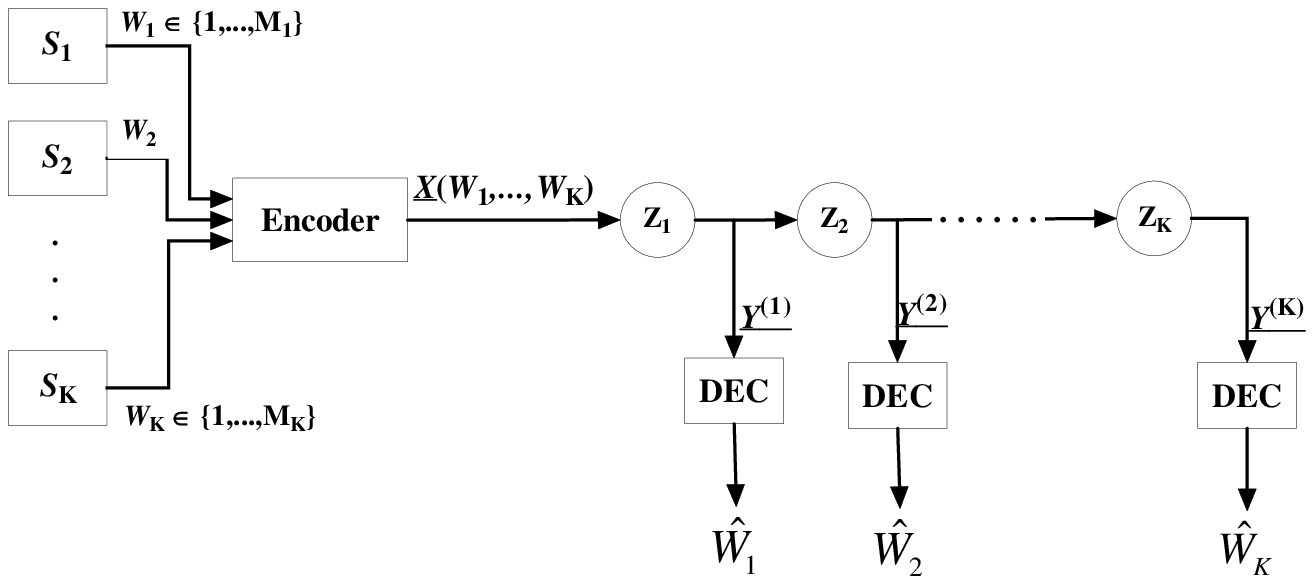}
  \caption{The communication system for a $K$-receiver broadcast Z channel.}\label{fig:sysKBZC}
\end{figure}

Theorem  \ref{theorem:capa_KBZC} below states that NE achieves the entire boundary of the capacity region for broadcast Z channels with any finite number of receivers. Consider the communication system for the $K$-receiver broadcast Z channel in Figure~\ref{fig:sysKBZC}. $\boldsymbol{X}=(X_1,\cdots,X_N)$ is a length-$N$ codeword determined by the messages $W_1,\cdots, W_K$. $\boldsymbol{Y}^{(1)}, \cdots,\boldsymbol{Y}^{(K)}$ are the channel
outputs corresponding to the channel input $\boldsymbol{X}$.

\begin{theorem}\label{theorem:capa_KBZC}
If $\sum_{i=1}^{N} \text{Pr}\{X_{i} = 0\}/N = q$, then no point
$(R_1,\cdots,R_K)$ such that
\begin{equation}
\begin{cases} R_j \geq \frac{q}{t_j}h(\beta_{j}t_j) -
\frac{q}{t_{j-1}}h(\beta_{j}t_{j-1}),\quad j=1,\cdots,K \\
R_d = \frac{q}{t_d}h(\beta_{d}t_d) -
\frac{q}{t_{d-1}}h(\beta_{d}t_{d-1})+ \delta, \quad \textrm{for some
} d \in \{1,\cdots, K\},\delta >0 \end{cases} \label{eq:rates}
\end{equation}
is achievable, where the $t_j$ are as in (\ref{eq:AchReg_KBZC}) and
(\ref{eq:t_define}).
\end{theorem}

Theorem \ref{theorem:capa_KBZC} indicates that no rate point
$(R_1,\cdots,R_K)$ outside the achievable region of the NE scheme is
achievable because if there
exists an achievable rate point $(R_1,\cdots,R_K)$ outside the NE
scheme's achievable region determined by (\ref{eq:AchReg_KBZC}),
then there must exist a boundary point $(R^*_1,\cdots,R^*_K)$ on the
NE scheme's achievable region such that $R_j \geq R^*_j$ for all
$j=1,\cdots,K$, and $R_d > R^*_d$ for some $d \in \{1,\cdots, K\}$.

The proof of Theorem \ref{theorem:capa_KBZC} uses the same basic approach as the proof of the converse of the coding theorem for Gaussian BCs \cite{Bergmans1973}. Lemma \ref{theorem:vectorZ} below plays the same role in this proof as the entropy power inequality does in the proof for Gaussian BCs.  We state and prove Lemma \ref{theorem:vectorZ} and then proceed with the proof of Theorem \ref{theorem:capa_KBZC}.

\begin{lemma}\label{theorem:vectorZ}
Consider the Markov chain $U \rightarrow \boldsymbol{X} \rightarrow
\boldsymbol{Y} \rightarrow \boldsymbol{Z}$ with $\sum_{i=1}^{N}
\text{Pr}(X_i = 0)/N = q$, if
\begin{equation}
H(\boldsymbol{Y}|U) \geq N \cdot \frac{q}{p} \cdot h(\beta_1 p),
\end{equation}
for some $p \in [q,1]$, then
\begin{align}
H(\boldsymbol{Z}|U) & \geq N \cdot \frac{q}{p} \cdot h(\beta_2 p)\\
& = N \cdot \frac{q}{p} \cdot h(\beta_1 p \beta_{\Delta}).
\end{align}
\end{lemma}

\begin{proof}[Proof of Lemma \ref{theorem:vectorZ}] Lemma
\ref{theorem:vectorZ} is the consequence of Proposition
\ref{theorem:vectorF} for the broadcast Z channel. Since
$H(\boldsymbol{Y}|U) \geq N \cdot q/p \cdot h(\beta_1 p)$,
\begin{align}
H(\boldsymbol{Z}|U) & \geq F^*_{T_{YX}^{(N)},T_{ZX}^{(N)}} (q, N
\cdot q/p \cdot h(\beta_1 p)) \label{eq:vectorZ_1}\\
& = N \cdot  F^*_{T_{YX},T_{ZX}} (q, q/p \cdot h(\beta_1 p))
 \label{eq:vectorZ_2}\\
& = N \cdot \frac{q}{p} \cdot h(\beta_2 p) \label{eq:vectorZ_3}\\
& = N \cdot \frac{q}{p} \cdot h(\beta_1 p
\beta_{\Delta}).\label{eq:vectorZ_4}
\end{align}
These steps are justified as follows:
\begin{itemize}
\item (\ref{eq:vectorZ_1}) follows from the definition of $F^*_{T_{YX}^{(N)},T_{ZX}^{(N)}} (\boldsymbol{q}, s)$;
\item (\ref{eq:vectorZ_2}) follows from Proposition \ref{theorem:vectorF};
\item (\ref{eq:vectorZ_3}) follows from the expression of the function $F^*$
for the broadcast Z channel in (\ref{eq:F_Z});
\item (\ref{eq:vectorZ_4}) follows from $\beta_{\Delta} = \text{Pr}\{Z=0|Y=0\}= \beta_2/\beta_1 $.
\end{itemize}
\end{proof}

\begin{proof}[Proof of Theorem \ref{theorem:capa_KBZC}] \replaced{The proof is by contradiction. To this end, suppose}{(by contradiction) Suppose}
 that the rates of (\ref{eq:rates}) are achievable, which
means that the probability of decoding error for each receiver can
be upper bounded by an arbitrarily small $\epsilon$ for sufficiently
large $N$
\begin{equation}
\textrm{Pr} \{ \hat{W} _{j} \neq W_j | \boldsymbol{Y}^{(j)} \} <
\epsilon , \quad j=1,\cdots,K.
\end{equation}
By Fano's inequality, this implies that
\begin{equation}
H(W_j|\boldsymbol{Y}^{(j)}) \leq h(\epsilon) + \epsilon \ln(M_j-1),
\quad j=1,\cdots,K. \label{eq:fano}
\end{equation}
Let $o(\epsilon)$ represent any function of $\epsilon$ such that
$o(\epsilon) \geq 0$ and $o(\epsilon) \rightarrow 0$ as $\epsilon
\rightarrow 0$. Equation (\ref{eq:fano}) implies that
$H(W_j|\boldsymbol{Y}^{(j)})$, $j=1,\cdots,K$, are all
$o(\epsilon)$. Therefore,
\begin{align}
H(W_j) & = H(W_j | W_{j+1}, \cdots, W_{K}) \label{eq:capa_KBZC_1}\\
& = I(W_j;\boldsymbol{Y}^{(j)} | W_{j+1}, \cdots, W_{K}) +
H(W_j|\boldsymbol{Y}^{(j)},W_{j+1}, \cdots, W_{K}) \\
& \leq I(W_j;\boldsymbol{Y}^{(j)} | W_{j+1}, \cdots, W_{K}) +
H(W_j|\boldsymbol{Y}^{(j)}) \\
& = H(\boldsymbol{Y}^{(j)} | W_{j+1}, \cdots, W_{K}) -
H(\boldsymbol{Y}^{(j)} | W_{j},W_{j+1}, \cdots, W_{K}) +
o(\epsilon), \label{eq:capa_KBZC_2}
\end{align}
where (\ref{eq:capa_KBZC_1}) follows from the independence of the
$W_j$, $j=1,\cdots,K$. From (\ref{eq:rates}), (\ref{eq:capa_KBZC_2})
and the fact that $N R_j \leq H(W_j)$,
\begin{equation}
H(\boldsymbol{Y}^{(j)} | W_{j+1}, \cdots, W_{K}) -
H(\boldsymbol{Y}^{(j)} | W_{j},W_{j+1}, \cdots, W_{K}) \geq N
\frac{q}{t_j}h(\beta_{j}t_j) - N
\frac{q}{t_{j-1}}h(\beta_{j}t_{j-1}) - o(\epsilon).
\label{eq:capa_KBZC_3}
\end{equation}
Next, using Lemma \ref{theorem:vectorZ} and (\ref{eq:capa_KBZC_3}),
we show in \deleted{the} Appendix \ref{app:B} that
\begin{equation}
H(\boldsymbol{Y}^{(K)}) \geq N h(\beta_{K}q) + N\delta -
o(\epsilon), \label{eq:contradiction}
\end{equation}
where $q=t_K = \sum_{i=1}^{N}\textrm{Pr}(X_i=0)/N$. Since $\epsilon$
can be arbitrarily small for sufficient large $N$, $o(\epsilon)
\rightarrow 0$ as $N \rightarrow \infty$. For sufficiently large
$N$, $H(\boldsymbol{Y}^{(K)}) \geq N h(\beta_{K}q) + N\delta /2$.
However, this contradicts
\begin{align}
H(\boldsymbol{Y}^{(K)}) & \leq \sum_{i=1}^{N}H(Y^{(K)}_{i}) \label{eq:capa_KBZC_5}\\
& = \sum_{i=1}^{N} h \left(\beta_{K} \cdot \textrm{Pr}(X_i=0) \right) \label{eq:capa_KBZC_6}\\
& \leq N  h \left(\beta_{K} \cdot \sum_{i=1}^{N}\textrm{Pr}(X_i=0)/N \right) \label{eq:capa_KBZC_7}\\
& = N h(\beta_{K}q).\label{eq:capa_KBZC_8}
\end{align}
Some of these steps are justified as follows:
\begin{itemize}
\item (\ref{eq:capa_KBZC_5}) follows from $\boldsymbol{Y}^{(K)} =
(Y^{(K)}_{1},\cdots,Y^{(K)}_{N})$;
\item (\ref{eq:capa_KBZC_7}) is obtained by applying Jensen's inequality to the concave function
$h(\cdot)$;
\item (\ref{eq:capa_KBZC_8}) follows from $q =
\sum_{i=1}^{N}\textrm{Pr}(X_i=0)/N$.
\end{itemize}

The desired contradiction has been obtained, so the theorem is
proved.
\end{proof}

\section{Input-Symmetric Degraded Broadcast Channels}
\label{sec:inputsymmetric}

The input-symmetric channel was first introduced in
\cite{Witsenhausen1975} and studied further in \cite{DDIC}
\cite{Liu2008} \cite{Symmetricchannel}. The definition of the
input-symmetric channel is as follows: Let $\Phi_{n}$ denote the
symmetric group of permutations of $n$ objects by  $n \times n$
permutation matrices. An $n$-input $m$-output channel with
transition probability matrix $T_{m \times n}$ is input-symmetric if
the set
\begin{equation}
\mbox{$\mathcal{G}$}_{T} = \left\{G \in \Phi_n |\exists \Pi \in
\Phi_m, \text{ s.t. } TG=\Pi T \right\}
\end{equation}
is transitive, which means for any $i,j \in \{1, \cdots ,n\}$, there
exists a permutation matrix $G \in \mbox{$\mathcal{G}$}_{T}$ which
maps the $i$-th row to the $j$-th row \cite{Witsenhausen1975}. An
important property of input-symmetric channels is that the uniform
distribution achieves capacity.  We extend the definition of the input-symmetric channel to the
input-symmetric DBC as follows:
\begin{definition}
{\em (Input-Symmetric Degraded Broadcast Channel)} A discrete memoryless
DBC $X \rightarrow Y \rightarrow Z$ with $|\mbox{$\mathcal{X}$}| =
k$, $|\mbox{$\mathcal{Y}$}| = n$ and $|\mbox{$\mathcal{Z}$}| = m$ is
input-symmetric if the set $\mbox{$\mathcal{G}$}_{T_{YX},T_{ZX}}$ is transitive where
\begin{align}
\mbox{$\mathcal{G}$}_{T_{YX},T_{ZX}} & \overset{\Delta}{=}
\mbox{$\mathcal{G}$}_{T_{YX}} \cap \mbox{$\mathcal{G}$}_{T_{ZX}} \\
& = \left\{G \in \Phi_k |\exists \Pi_{YX} \in \Phi_n, \Pi_{ZX} \in
\Phi_m, \text{ s.t. } T_{YX}G=\Pi_{YX} T_{YX} , T_{ZX}G=\Pi_{ZX}
T_{ZX} \right\}\, .
\end{align}

\end{definition}

 Lemmas \ref{theorem:inputsymm_group} and  \ref{theorem:inputsymm_sum} below establish basic properties of $\mbox{$\mathcal{G}$}_{T_{YX},T_{ZX}}$.

\begin{lemma}\label{theorem:inputsymm_group}
$\mbox{$\mathcal{G}$}_{T_{YX},T_{ZX}}$ is a group under matrix
multiplication.
\end{lemma}

\begin{proof} Every closed subset of a group is a group. Since
$\mbox{$\mathcal{G}$}_{T_{YX},T_{ZX}}$ is a subset of $\Phi_k$,
which is a group under matrix multiplication, it suffices to show
that $\mbox{$\mathcal{G}$}_{T_{YX},T_{ZX}}$ is closed under matrix
multiplication. Suppose $G_1, G_2 \in
\mbox{$\mathcal{G}$}_{T_{YX},T_{ZX}}$ such that
$T_{YX}G_1=\Pi_{YX,1} T_{YX}$, $T_{ZX}G_1=\Pi_{ZX,1} T_{ZX}$,
$T_{YX}G_2=\Pi_{YX,2} T_{YX}$ and $T_{ZX}G_2=\Pi_{ZX,2} T_{ZX}$.
Thus,
\begin{equation}
T_{YX}G_1 G_2 = \Pi_{YX,1} \Pi_{YX,2} T_{YX},
\end{equation}
and
\begin{equation}
T_{ZX}G_1 G_2 = \Pi_{ZX,1} \Pi_{ZX,2} T_{ZX}.
\end{equation}
Therefore, $G_1 G_2 \in \mbox{$\mathcal{G}$}_{T_{YX},T_{ZX}}$.
\end{proof}

\begin{lemma}\label{theorem:inputsymm_sum}
Let $l = |\mbox{$\mathcal{G}$}_{T_{YX},T_{ZX}}|$ so that
$\mbox{$\mathcal{G}$}_{T_{YX},T_{ZX}} \overset{\Delta}{=}\mbox{$\mathcal{G}$}_{T_{YX}} \cap \mbox{$\mathcal{G}$}_{T_{ZX}} = \{ G_1, \cdots, G_l\}$.  Also let
$k = |\mathcal{X}|$. Then $\sum_{i=1}^{l} G_i = \frac{l}{k}
\boldsymbol{1}\boldsymbol{1}^T$, where $\frac{l}{k}$ is an integer
and $\boldsymbol{1}$ is an all-ones vector.
\end{lemma}

\begin{proof} For all $j = 1, \cdots, l$,
\begin{equation}
G_j \left(\sum_{i=1}^{l} G_i \right) \stackrel{(a)}{=}
\sum_{i=1}^{l} G_j G_i \stackrel{(b)}{=} \sum_{i=1}^{l} G_i,
\label{eq:inputsymm_sum}
\end{equation}
where (a) follows from the distributive law for the field of rational matrices and (b) follows from the closure axiom and the
inverse element axiom for the group
$\mbox{$\mathcal{G}$}_{T_{YX},T_{ZX}}$.

Hence, $\sum_{i=1}^{l} G_i$ has $k$ identical columns and $k$
identical rows since $\mbox{$\mathcal{G}$}_{T_{YX},T_{ZX}}$ is
transitive. Therefore, $\sum_{i=1}^{l} G_i = \frac{l}{k}
\boldsymbol{1}\boldsymbol{1}^T$.
\end{proof}

\begin{definition}
\added{{\em (Smallest Transitive Set)}} A subset of $\mbox{$\mathcal{G}$}_{T_{YX},T_{ZX}}$,
$\{G_{i_1},\cdots , G_{i_{l_s}}\}$, is a smallest transitive subset
of $\mbox{$\mathcal{G}$}_{T_{YX},T_{ZX}}$ if
\begin{equation}
\sum_{j=1}^{l_s} G_{i_j} = \frac{l_s}{k}
\boldsymbol{1}\boldsymbol{1}^T, \label{eq:smallest}
\end{equation}
where $\frac{l_s}{k}$ is the smallest possible integer for which
(\ref{eq:smallest}) is satisfied.
\end{definition}

\subsection{Examples: binary-symmetric BCs and binary-erasure BCs}
The class of input-symmetric DBCs includes most of the common
discrete memoryless DBCs. For example, the binary-symmetric BC $X
\rightarrow Y \rightarrow Z$ with marginal transition probability
matrices
\begin{equation}
T_{YX}=\begin{bmatrix}
1- \alpha_1 & \alpha_1\\
\alpha_1 & 1- \alpha_1
\end{bmatrix} \text{ and }
T_{ZX}=\begin{bmatrix}
1- \alpha_2 & \alpha_2\\
\alpha_2 & 1- \alpha_2
\end{bmatrix}, \nonumber
\end{equation}
where $0 \leq \alpha_1 \leq \alpha_2 \leq 1/2$, is input-symmetric
since
\begin{equation} \label{G_BBEC}
\mbox{$\mathcal{G}$}_{T_{YX},T_{ZX}} = \left\{\begin{bmatrix}
1 & 0\\
0 & 1
\end{bmatrix} ,
\begin{bmatrix}
0 & 1 \\
1 & 0
\end{bmatrix} \right\}
\end{equation}
is transitive.

Another interesting example is the binary-erasure BC with marginal
transition probability matrices
\begin{equation}
T_{YX}=\begin{bmatrix}
1- a_1 & 0\\
a_1 & a_1 \\
 0 & 1- a_1
\end{bmatrix} \text{ and }
T_{ZX}=\begin{bmatrix}
1- a_2 & 0\\
a_2 & a_2 \\
 0 & 1- a_2
\end{bmatrix},\nonumber
\end{equation}
where $0 \leq a_1 \leq a_2 \leq 1$. It is input-symmetric since its
$\mbox{$\mathcal{G}$}_{T_{YX},T_{ZX}}$ is the same as that of the
binary-symmetric BC shown in (\ref{G_BBEC}).

\subsection{Group-\replaced{Operation}{additive} DBCs are input-symmetric.}

\begin{figure}
  \centering
  \includegraphics[width=0.40\textwidth]{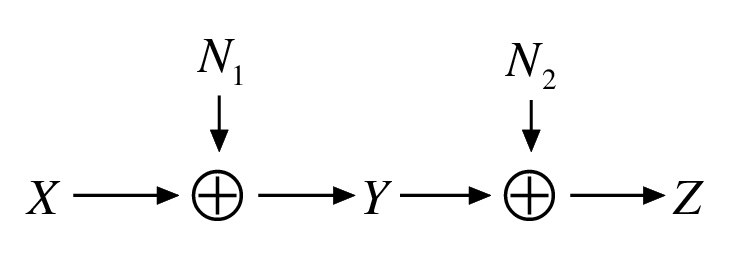}
  \caption{The group-\replaced{operation}{additive} degraded broadcast channel.}\label{fig:groupadditionchannel}
\end{figure}

We now define group-\replaced{operation}{additive}  DBCs and show that they are input symmetric.
\begin{definition}
{\em (Group-\replaced{Operation}{additive}  Degraded Broadcast
Channel)} A discrete DBC $X \rightarrow Y \rightarrow Z$ with
$\mathcal{X,Y,Z} $ $=\{1,\cdots, n\}$ is a
group-\replaced{operation}{additive} DBC if there exist two $n$-ary
random variables $N_1$ and $N_2$ such that $Y \sim X \oplus N_1$ and
$Z \sim Y \oplus N_2$ as shown in
Figure~\ref{fig:groupadditionchannel}, where $\sim$ denotes
identical distribution and $\oplus$ denotes a group
\replaced{operation}{addition} which is an operation that satisfies
the group axioms on the set $\{1,\cdots, n\}$.
\end{definition}

Group-\replaced{operation}{additive} DBCs include the binary-symmetric BC and the
discrete additive DBC of \cite{Benzel1979} as special cases. It is also
a channel model for Gaussian broadcast communication systems with
phase-shift-keying (PSK) modulation at the transmitter and direct
hard decisions on modulated symbols at the receivers.

\begin{theorem}\label{theorem:GroupAdditionInputSymmetric}
Group-\replaced{operation}{additive} DBCs are input-symmetric.
\end{theorem}

\begin{proof} For the  group-\replaced{operation}{additive} DBC $X \rightarrow Y
\rightarrow Z$ with $\mathcal{X,Y,Z} $ $=\{1,\cdots, n\}$, let
$G_{x}$ for $x= 1,\cdots,n$, be 0-1 matrices with entries
\begin{equation}
G_{x}(i,j)=
\begin{cases}
1 &\text{if $j \oplus x = i$}\\
0&\text{otherwise}
\end{cases}\, \text{for } i,j=1,\cdots,n.
\end{equation}
$G_{x}$ for $x= 1,\cdots,n$, are actually permutation matrices and
have the property that $G_{x_1} \cdot G_{x_2} = G_{x_2} \cdot
G_{x_1} = G_{x_1 \oplus x_2}$. Let $[\gamma_1, \cdots,
\gamma_{n}]^T$ be the distribution of $N_1$. Since $Y$ has the same
distribution as $X \oplus N_1$, one has
\begin{equation}
T_{YX}= \sum_{x=1}^{n} \gamma_{x} G_{x}.
\end{equation}
Hence, $T_{YX} G_{x} = G_{x} T_{YX}$ for all $x=1,\cdots,n$.
Similarly, we have $T_{ZX} G_{x} = G_{x} T_{ZX}$ for all
$x=1,\cdots,n$, and so
\begin{equation}
\{G_{1},\cdots,G_{n}\} \subseteq
\mbox{$\mathcal{G}$}_{T_{YX},T_{ZX}}.
\end{equation}
Since the set $\{G_{1},\cdots,G_{n}\}$ is transitive by definition,
$\mbox{$\mathcal{G}$}_{T_{YX},T_{ZX}}$ is also transitive and hence
the group-\replaced{operation}{additive} DBC is input-symmetric.
\end{proof}

By definition, $\sum_{j=1}^{n} G_j =
\boldsymbol{1}\boldsymbol{1}^T$, and hence, $\{G_{1},\cdots,G_{n}\}$
is a smallest transitive subset of
$\mbox{$\mathcal{G}$}_{T_{YX},T_{ZX}}$ for the group-\replaced{operation}{additive} DBC.

\subsection{A note on discrete degraded interference channels (DDICs)} We briefly note that while DDICs and their related DBCs are closely related to IS-DBCs, the class of IS-DBCs is not addressed by  \cite{DDIC} or \cite{Liu2008}.  The class of DDICs and the corresponding DBCs studied in \cite{DDIC}
and \cite{Liu2008} have to satisfy the condition that the transition
probability matrix $T_{ZY}$ is input-symmetric, i.e.,
$\mbox{$\mathcal{G}$}_{T_{ZY}}$ is transitive. The input-symmetric
DBC, however, does not have to satisfy this condition. The following
example provides an IS-DBC which is not covered in \cite{DDIC}
\cite{Liu2008}. Consider a binary-input DBC $X \rightarrow Y \rightarrow Z$ with
transition probability matrices
\begin{equation}
T_{YX}=\begin{bmatrix}
a & c\\
b & d\\
c & a\\
d & b
\end{bmatrix},
T_{ZY}=\begin{bmatrix}
e & f & g & h\\
g & h & e & f
\end{bmatrix},\nonumber
\end{equation}
and
\begin{equation}
T_{ZX} = T_{ZY}T_{YX} = \begin{bmatrix}
\alpha & \beta \\
\beta & \alpha\\
\end{bmatrix},
\end{equation}
where $a+c = b+d = 1$, $e+f+g+h=1$, $\alpha = ae+bf+cg+dh$ and
$\beta = ag + bh+ce+df$. This DBC is input-symmetric since its
$\mbox{$\mathcal{G}$}_{T_{YX},T_{ZX}}$ is the same as that of the
 broadcast binary-symmetric channel shown in (\ref{G_BBEC}). It is
not covered by the results of \cite{DDIC} \cite{Liu2008} because
\begin{equation}
\mbox{$\mathcal{G}$}_{T_{ZY}} = \left\{\begin{bmatrix}
1 & 0 & 0 & 0\\
0 & 1 & 0 & 0\\
0 & 0 & 1 & 0\\
0 & 0 & 0 & 1
\end{bmatrix} ,
\begin{bmatrix}
0 & 0 & 1 & 0 \\
0 & 0 & 0 & 1 \\
1 & 0 & 0 & 0 \\
0 & 1 & 0 & 0
\end{bmatrix} \right\}
\end{equation}
is \emph{not} transitive.

\subsection{Optimal input distribution and capacity region for IS-DBCs}
Consider the input-symmetric DBC $X \rightarrow Y \rightarrow Z$
with the marginal transition probability matrices $T_{YX}$ and
$T_{ZX}$. Recall that the set $\mathcal{C}$ is the set of all
$(\boldsymbol{p}_{X},s,\eta)$ satisfying (\ref{eq:p}), (\ref{eq:xi})
and (\ref{eq:eta}) for some choice of $l$, $\boldsymbol{w}$ and
$T_{XU}$,
the set $\mathcal{C}^*$ $ = \{ (s , \eta ) |
(\boldsymbol{p}_{X},s,\eta) \in $ $\mathcal{C}$ $ \text{ for some }
\boldsymbol{p}_{X} \} $ is the projection of the set $\mathcal{C}$
on the $(s , \eta)$-plane, and the set
$\mathcal{C}^*_{\boldsymbol{q}}$ is the subset of
$\mathcal{C}^*$ for which $\boldsymbol{p}_{X} = \boldsymbol{q}$.

\begin{lemma}\label{theorem:p2Gp}
For any permutation matrix $G \in
\mbox{$\mathcal{G}$}_{T_{YX},T_{ZX}}$ and $(\boldsymbol{p},s,\eta)
\in \mathcal{C}$,  $(G \boldsymbol{p},s,\eta) \in
\mathcal{C}$.
\end{lemma}

\begin{proof} Since $(\boldsymbol{p},s,\eta)$
satisfies (\ref{eq:p}), (\ref{eq:xi}) and
(\ref{eq:eta}) for some choice of $l$, $\boldsymbol{w}$ and
 $T_{XU}= [\boldsymbol{t}_1 \cdots
\boldsymbol{t}_l]$,

\begin{align}
G T_{XU} \boldsymbol{w}& = G \boldsymbol{p} \\
\sum _{j =1}^{l}w_{j} h_{n}(T_{YX} G \boldsymbol{t}_{j}) & = \sum
_{j =1}^{l}w_{j} h_{n}(\Pi_{YX} T_{YX}
 \boldsymbol{t}_{j}) = s \\
\sum _{j =1}^{l}w_{j} h_{m}(T_{ZX} G \boldsymbol{t}_{j}) & = \sum
_{j =1}^{l}w_{j} h_{m}( \Pi_{ZX} T_{ZX} \boldsymbol{t}_{j}) = \eta.
\end{align}Hence, $( G \boldsymbol{p},s,\eta)$ satisfies (\ref{eq:p}),
(\ref{eq:xi}) and (\ref{eq:eta}) for the choice of $l$,
$\boldsymbol{w}$ and $G T_{XU} $.
\end{proof}

\begin{corollary}\label{theorem:Cp2CGp}
$\forall \boldsymbol{p} \in \Delta_{k}$ and $G \in
\mbox{$\mathcal{G}$}_{T_{YX},T_{ZX}}$, one has
$\mbox{$\mathcal{C}$}^*_{G\boldsymbol{p}} =
\mbox{$\mathcal{C}$}^*_{\boldsymbol{p}} $, and so
$F^*(G\boldsymbol{p},s) = F^*(\boldsymbol{p},s)$ for any $H(Y|X)
\leq s \leq H(Y)$.
\end{corollary}

\begin{lemma}\label{theorem:Cstar_Cu}
For any input-symmetric DBC, $\mbox{$\mathcal{C}$}^{*} =
\mbox{$\mathcal{C}$}^*_{\boldsymbol{u}}$, where $\boldsymbol{u}$
denotes the uniform distribution.
\end{lemma}

\begin{proof} For any $(s,\eta) \in \mbox{$\mathcal{C}$}^{*}$, there
exits a distribution $\boldsymbol{p}$ such that
$(\boldsymbol{p},s,\eta) \in \mbox{$\mathcal{C}$}$. Let
$\mbox{$\mathcal{G}$}_{T_{YX},T_{ZX}} = \{G_1,\cdots,G_l\}$. By
Corollary \ref{theorem:Cp2CGp}, $( G_j \boldsymbol{p},s,\eta) \in
\mbox{$\mathcal{C}$}$ for all $j=1,\cdots,l$.  By the convexity of
the set $\mathcal{C}$,
\begin{equation}
(\boldsymbol{q},s,\eta) = \left(  \sum_{j=1}^{l}
 \frac{1}{l} G_j \boldsymbol{p} \; , s , \eta \right) \in \mathcal{C},
\end{equation}
where $\boldsymbol{q} = \sum_{j=1}^l \frac{1}{l} G_j
\boldsymbol{p}$. Since $\mbox{$\mathcal{G}$}_{T_{YX},T_{ZX}}$ is a
group
, for any permutation matrix $G' \in
\mbox{$\mathcal{G}$}_{T_{YX},T_{ZX}}$,
\begin{equation}
G' \boldsymbol{q} \; = \! \sum_{j=1}^l \!\! \frac{1}{l} G' G_j
\boldsymbol{p} \; = \! \sum_{j=1}^l \!\! \frac{1}{l} G_j
\boldsymbol{p} \; =\; \boldsymbol{q}.
\end{equation}
Since $G' \boldsymbol{q} = \boldsymbol{q}$, the $i^{\text{th}}$
entry and the $j^{\text{th}}$ entry of $\boldsymbol{q}$ are the same
if $G'$ permutes the $i^{\text{th}}$ row to the $j^{\text{th}}$ row.
Since the set $\mbox{$\mathcal{G}$}_{T_{YX},T_{ZX}}$ for an
input-symmetric DBC is transitive, all the entries of
$\boldsymbol{q}$ are the same, and so $\boldsymbol{q} =
\boldsymbol{u}$. This implies that $(s,\eta) \in
\mbox{$\mathcal{C}$}^*_{\boldsymbol{u}}$. Since $(s,\eta)$ is
arbitrarily taken from $\mbox{$\mathcal{C}$}^{*}$, one has
$\mbox{$\mathcal{C}$}^{*} \subseteq
\mbox{$\mathcal{C}$}^*_{\boldsymbol{u}}$. On the other hand, by
definition, $\mbox{$\mathcal{C}$}^{*} \supseteq
\mbox{$\mathcal{C}$}^*_{\boldsymbol{u}}$. Therefore,
$\mbox{$\mathcal{C}$}^{*} =
\mbox{$\mathcal{C}$}^*_{\boldsymbol{u}}$.
\end{proof}

Now we state and prove that the uniformly distributed $X$ is optimal
for input-symmetric DBCs.

\begin{theorem}\label{theorem:unif_symm}
For any input-symmetric DBC, its capacity region can be achieved by
using the transmission strategies such that the broadcast signal $X$
is uniformly distributed. As a consequence, the capacity region is
\begin{equation}\label{eq:capa_symm} \bar{\text{co}}\left \{
(R_1,R_2):R_1 \leq s - h_{n}(T_{YX}\boldsymbol{e_1}), R_2 \leq
h_{m}(T_{ZX}\boldsymbol{u})- F^*_{T_{YX},T_{ZX}}(\boldsymbol{u},s),
h_{n}(T_{YX}\boldsymbol{e_1}) \leq s \leq \ln(n) \right\},
\end{equation}
where $\boldsymbol{e_1} = [1,0,\cdots,0]^T$, $n = | \mathcal{Y}|$,
and $m = |\mathcal{Z}|$.
\end{theorem}

\begin{proof} Let $\boldsymbol{q} = [q_1, \cdots,q_k]^T$ be the
distribution of the channel input $X$ for the input-symmetric DBC $X
\rightarrow Y \rightarrow Z$. Since $\mbox{$\mathcal{G}$}_{T_{YX}}$
is transitive, the columns of $T_{YX}$ are permutations of each
other.
\begin{align}
H(Y|X) & = \sum_{i=1}^{k} q_{i} H(Y|X=i)\\
 & = \sum_{i=1}^{k} q_i h_{n}(T_{YX} \boldsymbol{e_i})\\
 & = \sum_{i=1}^{k} q_i h_{n}(T_{YX} \boldsymbol{e_1})\\
 & = h_{n}(T_{YX} \boldsymbol{e_1}) \label{eq:unif_symm_1},
\end{align}
which is independent of $\boldsymbol{q}$. Let $l=
|\mbox{$\mathcal{G}$}_{T_{YX},T_{ZX}}|$ and
$\mbox{$\mathcal{G}$}_{T_{YX},T_{ZX}} = \{G_1,\cdots,G_l\}$.
\begin{align}
H(Z) & = h_{m}(T_{ZX}\boldsymbol{q}) \\
& = \frac{1}{l}\sum_{i=1}^{l} h_{m}(T_{ZX}G_{i}\boldsymbol{q}) \\
& \leq h_{m}\left(T_{ZX} \sum_{i=1}^{l} \frac{1}{l} G_{i}\boldsymbol{q} \right) \label{eq:unif_symm_11}\\
& = h_{m}(T_{ZX}\boldsymbol{u}), \label{eq:unif_symm_2}
\end{align}
where (\ref{eq:unif_symm_11}) follows from Jensen's inequality.
Since $\mbox{$\mathcal{C}$}^{*} =
\mbox{$\mathcal{C}$}^*_{\boldsymbol{u}}$ for the input-symmetric
DBC,
\begin{equation}
F^*(\boldsymbol{q},s) \geq F^*(\boldsymbol{u},s).
\label{eq:unif_symm_3}
\end{equation}
Plugging (\ref{eq:unif_symm_1}), (\ref{eq:unif_symm_2}) and
(\ref{eq:unif_symm_3}) into (\ref{eq:MyCapa3}), the expression of
the capacity region for the DBC, the capacity region for
input-symmetric DBCs is
\begin{align}
& \bar{\text{co}}\left [ \bigcup _{\boldsymbol{p}_{X}=\boldsymbol{q}
\in \Delta_{k}} \left \{ (R_1,R_2):R_1 \leq s - H(Y|X), R_2 \leq
H(Z)- F^*_{T_{YX},T_{ZX}}(\boldsymbol{q},s) \right\} \right ] \label{eq:IS_capa1} \\
\subseteq \; & \bar{\text{co}}\left [ \bigcup
_{\boldsymbol{p}_{X}=\boldsymbol{q} \in \Delta_{k}} \left \{
(R_1,R_2):R_1 \leq s - h_{n}(T_{YX} \boldsymbol{e_1}), R_2 \leq
h_{m}(T_{ZX}\boldsymbol{u})-
F^*_{T_{YX},T_{ZX}}(\boldsymbol{u},s) \right\} \right ] \label{eq:IS_capa2} \\
= \; & \bar{\text{co}}\left \{ (R_1,R_2):R_1 \leq s -
h_{n}(T_{YX}\boldsymbol{e_1}), R_2 \leq h_{m}(T_{ZX}\boldsymbol{u})-
F^*_{T_{YX},T_{ZX}}(\boldsymbol{u},s)\right\}\label{eq:IS_capa3} \\
= \; & \bar{\text{co}} \left \{ (R_1,R_2): \boldsymbol{p}_{X}=
\boldsymbol{u}, R_1 \leq s - H(Y|X), R_2 \leq H(Z)-
F^*_{T_{YX},T_{ZX}}(\boldsymbol{u},s) \right\}
\label{eq:IS_capa4} \\
\subseteq & \bar{\text{co}}\left [ \bigcup
_{\boldsymbol{p}_{X}=\boldsymbol{q} \in \Delta_{k}} \left \{
(R_1,R_2):R_1 \leq s - H(Y|X), R_2 \leq H(Z)-
F^*_{T_{YX},T_{ZX}}(\boldsymbol{q},s) \right\} \right ],
\label{eq:IS_capa5}
\end{align}

Note that (\ref{eq:IS_capa1}) and (\ref{eq:IS_capa5}) are identical
expressions, hence (\ref{eq:IS_capa1} - \ref{eq:IS_capa5}) are all
equal. Therefore, (\ref{eq:capa_symm}) and (\ref{eq:IS_capa3})
express the capacity region for the input-symmetric DBC, which also
means that the capacity region can be achieved by using transmission
strategies where the broadcast signal $X$ is uniformly distributed.
\end{proof}

\subsection{Permutation encoding approach and its optimality for IS-DBCs}

\begin{figure}
  \centering
  \includegraphics[width=0.80\textwidth]{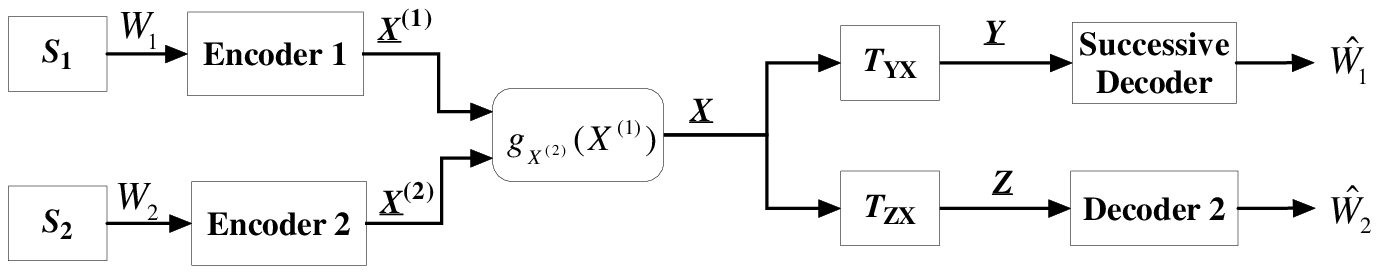}
  \caption{The block diagram of the permutation encoding approach.}\label{fig:permapproach}
\end{figure}

The permutation encoding approach is an independent-encoding scheme
which achieves the capacity region for input-symmetric DBCs. The
block diagram of this approach is shown in
Figure~\ref{fig:permapproach}. In Figure~\ref{fig:permapproach}, $W_1$
is the message for Receiver 1, which sees the less-degraded channel
$T_{YX}$, and $W_2$ is the message for Receiver 2, which sees the
more-degraded channel $T_{ZX}$. The permutation encoding approach is
first to independently encode these two messages into two codewords
$\boldsymbol{X}^{(1)}$ and $\boldsymbol{X}^{(2)}$, and then to
combine these two independent codewords using a single-letter
operation.

Let $\mbox{$\mathcal{G}$}_{s}$ be a smallest transitive subset of
$\mbox{$\mathcal{G}$}_{T_{YX},T_{ZX}}$. Denote $k = |\mathcal{X}|$
and $l_s= |\mbox{$\mathcal{G}$}_s|$. Use a random coding technique
to design the codebook for Receiver 1 according to the $k$-ary
random variable $X^{(1)}$ with distribution $\boldsymbol{p_1}$ and
the codebook for Receiver 2 according to the $l_s$-ary random
variable $X^{(2)}$ with uniform distribution. Let
$\mbox{$\mathcal{G}$}_{s}= \{G_1,\cdots,G_{l_s} \}$. Define the
permutation function $g_{x^{(2)}}(x^{(1)}) = x$ if the permutation
matrix $G_{x^{(2)}}$ maps the $x^{(1)}$-th column to the $x$-th
column, where $x^{(2)} \in \{1,\cdots,l_s\}$ and $x,x^{(1)} \in \{1,
\cdots, k\}$. Hence, $g_{x^{(2)}}(x^{(1)}) = x$ if and only if the
$x^{(1)}$-th row, $x$-th column entry of $G_{x^{(2)}}$ is 1. The
permutation encoding approach is then to broadcast $\boldsymbol{X}$
which is obtained by applying the single-letter permutation function
$X = g_{X^{(2)}}(X^{(1)})$ on symbols of codewords
$\boldsymbol{X}^{(1)}$ and $\boldsymbol{X}^{(2)}$. Since $X^{(2)}$
is uniformly distributed and $\sum_{j=1}^{l_s} G_j =
\frac{l_s}{k}\boldsymbol{1}\boldsymbol{1}^{T}$, the broadcast signal
$X$ is also uniformly distributed.

\begin{figure}
  \centering
  \includegraphics[width=0.50\textwidth]{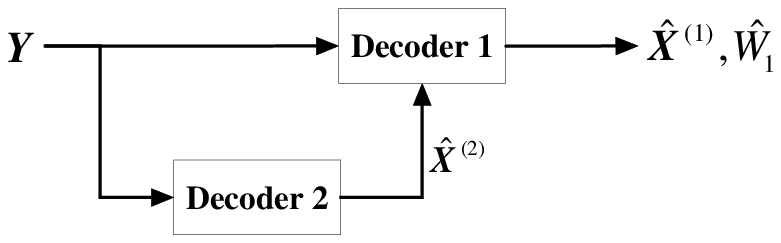}
  \caption{The structure of the successive decoder for input-symmetric DBCs.}\label{fig:successivedecoder}
\end{figure}

Receiver 2 receives $\boldsymbol{Z}$ and decodes the desired message
directly. Receiver 1 receives $\boldsymbol{Y}$ and successively
decodes the message for Receiver 2 and then for Receiver 1. The
structure of the successive decoder is shown in
Figure~\ref{fig:successivedecoder}. Note that Decoder 1 in
Figure~\ref{fig:successivedecoder} is \emph{not} a joint decoder even
though it has two inputs $\boldsymbol{Y}$ and
$\boldsymbol{\hat{X}}^{(2)}$.

\begin{figure}
  \centering
  \includegraphics[width=0.50\textwidth]{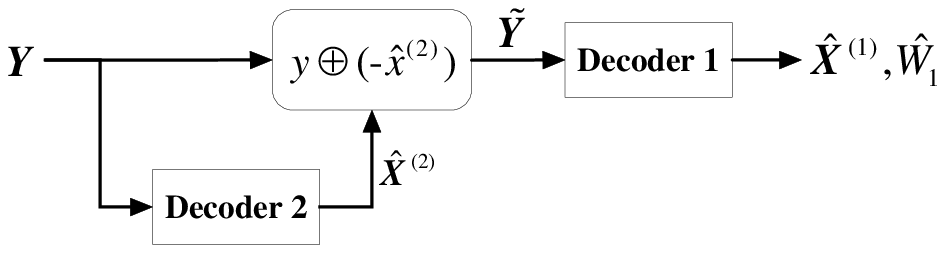}
  \caption{The structure of the successive decoder for degraded group-\replaced{operation}{addition} DBCs.}\label{fig:groupadd_decoder}
\end{figure}

In particular, for the group-\replaced{operation}{additive} DBC with $Y \sim X \oplus N_1$
and $Z \sim Y \oplus N_2$, the permutation function
$g_{x^{(2)}}(x^{(1)})$ is the group \replaced{operation}{addition} $x^{(2)} \oplus
x^{(1)}$. Hence the permutation encoding approach for the
group-\replaced{operation}{additive} DBC is the NE scheme for the group-\replaced{operation}{additive} DBC. The
successive decoder for the group-\replaced{operation}{additive} DBC is shown in
Figure~\ref{fig:groupadd_decoder}, where
\begin{equation}
\tilde{y} = y \oplus (-\hat{x}^{(2)}).
\end{equation}

From the analysis of successive decoding in the proof of the coding theorem for DBCs \cite{Bergmans1973}
\cite{Gallager1974}, the achievable region of the permutation
encoding approach for the input-symmetric DBC is determined by
\begin{align}
R_1 & \leq I(X;Y|X^{(2)}) \label{eq:capa_symm_1}\\
& = H(Y|X^{(2)}) - H(Y|X) \label{eq:capa_symm_2}\\
& = \sum_{x^{(2)} = 1}^{l_s}\text{Pr}(X^{(2)}=x^{(2)})H(Y|X^{(2)}=x^{(2)}) - \sum_{x=1}^{k}\text{Pr}(X=x)H(Y|X=x) \label{eq:capa_symm_3}\\
& = \sum_{x^{(2)} = 1}^{l_s}\text{Pr}(X^{(2)}=x^{(2)})h_{n}(T_{YX}G_{x^{(2)}}\boldsymbol{p_1}) - \sum_{x=1}^{k}\text{Pr}(X=x)h_{n}(T_{YX}\boldsymbol{e}_x) \label{eq:capa_symm_4}\\
& = \sum_{x^{(2)} = 1}^{l_s}\text{Pr}(X^{(2)}=x^{(2)})h_{n}(\Pi_{YX,x^{(2)}}T_{YX}\boldsymbol{p_1}) - \sum_{x=1}^{k}\text{Pr}(X=x)h_{n}(T_{YX}\boldsymbol{e}_1) \label{eq:capa_symm_5}\\
& = h_{n}(T_{YX}\boldsymbol{p_1}) - h_{n}(T_{YX}\boldsymbol{e}_1),
\label{eq:capa_symm_6}
\end{align}
and
\begin{align}
R_2 & \leq I(X^{(2)};Z) \label{eq:capa_symm_7}\\
& = H(Z) - H(Z|X^{(2)}) \label{eq:capa_symm_8}\\
& = h_{m}(T_{ZX}\boldsymbol{u}) -
\sum_{x^{(2)}=1}^{l_s}\text{Pr}(X^{(2)}=x^{(2)})h_{m}(T_{ZX}G_{x^{(2)}}\boldsymbol{p_1}) \label{eq:capa_symm_9}\\
& = h_{m}(T_{ZX}\boldsymbol{u}) - \sum_{x^{(2)}=1}^{l_s}\text{Pr}(X^{(2)}=x^{(2)})h_{m}(\Pi_{ZX,x^{(2)}}T_{ZX}\boldsymbol{p_1}) \label{eq:capa_symm_10}\\
& = h_{m}(T_{ZX}\boldsymbol{u}) - h_{m}(T_{ZX}\boldsymbol{p_1})
\label{eq:capa_symm_11},
\end{align}
where $\boldsymbol{u}$ is the $k$-ary uniform distribution,
$\boldsymbol{p}_1$ is the distribution of $X^{(1)}$, and
$\boldsymbol{e}_x$ is a 0-1 vector such that the $x$-th entry is 1
and all other entries are 0. Hence, the achievable region is
\begin{equation} \label{eq:achievable_symm}
\bar{\text{co}} \left [ \bigcup _{\boldsymbol{p_1} \in \Delta_{k}}
\left \{ (R_1,R_2):R_1 \leq h_{n}(T_{YX}\boldsymbol{p_1}) -
h_{n}(T_{YX}\boldsymbol{e_1}), R_2 \leq h_{m}(T_{ZX}\boldsymbol{u})-
h_{m}(T_{ZX}\boldsymbol{p_1})\right\} \right ]
\end{equation}

Define $\tilde{F}(s)$ as the infimum of
$h_{m}(T_{ZX}\boldsymbol{p_1})$ with respect to all distributions
$\boldsymbol{p_1}$ such that $h_{n}(T_{YX}\boldsymbol{p_1}) = s$.
Hence the achievable region (\ref{eq:achievable_symm}) can be
expressed as
\begin{equation} \label{eq:achievable_symm2}
 \left \{ (R_1,R_2):R_1 \leq s -
h_{n}(T_{YX}\boldsymbol{e_1}), R_2 \leq h_{m}(T_{ZX}\boldsymbol{u})-
\underline{\text{env}}\tilde{F}(s), h_{n}(T_{YX}\boldsymbol{e_1})
\leq s \leq h_{n}(T_{YX}\boldsymbol{u}) \right\},
\end{equation}
where $\underline{\text{env}}\tilde{F}(s)$ denotes the lower convex
envelope of $\tilde{F}(s)$.

\begin{theorem}\label{theorem:opt_symm}
The permutation encoding approach achieves the capacity region for
input-symmetric DBCs, which is expressed in (\ref{eq:capa_symm}),
(\ref{eq:achievable_symm}) and (\ref{eq:achievable_symm2}).
\end{theorem}

\begin{proof} In order to show that the achievable region
(\ref{eq:achievable_symm2}) is the same as the capacity region
(\ref{eq:capa_symm}) for the input-symmetric DBC, it suffices to
show that
\begin{equation}
\underline{\text{env}}\tilde{F}(s) \leq F^*(\boldsymbol{u},s).
\end{equation}

For any $p(u,x)$ with \replaced{uniformly distributed $X$}{$\boldsymbol{p}_{X} = \boldsymbol{u}$},
\begin{align}
H(Z|U) & = \sum_{u} \text{Pr}(U=u)H(Z|U=u) \\
& = \sum_{u} \text{Pr}(U=u)h_{m}(T_{ZX}\boldsymbol{p}_{X|U=u}) \\
& \geq \sum_{u} \text{Pr}(U=u)\tilde{F}(h_{n}(T_{YX}\boldsymbol{p}_{X|U=u})) \label{eq:capa_symm_12}\\
& \geq \sum_{u} \text{Pr}(U=u) \underline{\text{env}} \tilde{F}\left ( h_{n}(T_{YX}\boldsymbol{p}_{X|U=u}) \right )\label{eq:capa_symm_13} \\
& \geq \underline{\text{env}} \tilde{F} \left ( \sum_{u}
\text{Pr}(U=u)h_{n}(T_{YX}\boldsymbol{p}_{X|U=u}) \right ) \label{eq:capa_symm_14}\\
& = \underline{\text{env}} \tilde{F}(H(Y|U)),\label{eq:capa_symm_15}
\end{align}
where $\boldsymbol{p}_{X|U=u}$ is the conditional distribution of
$X$ given $U=u$. Some of these steps are justified as follows:
\begin{itemize}
\item (\ref{eq:capa_symm_12}) follows from the definition of
$\tilde{F}(s)$;
\item (\ref{eq:capa_symm_14}) follows from Jensen's inequality.
\end{itemize}
Combining (\ref{eq:capa_symm_15}) and the definition of $F^*$, one
has $\underline{\text{env}}\tilde{F}(s) \leq F^*(\boldsymbol{u},s)$.
\end{proof}

\begin{corollary}\label{theorem:opt_groupadd}
The NE scheme achieves the capacity region for group-\replaced{operation}{additive} DBCs.
\end{corollary}

\begin{conjecture}\label{theorem:minsubset}
The alphabet size of the code for Receiver 2, $l_s$, is equal to the
alphabet size of the channel input, $k$, in a permutation encoding
approach for any input-symmetric DBC. In other words, a smallest
transitive subset $\{G_{1},\cdots , G_{l_s}\}$ of
$\mbox{$\mathcal{G}$}_{T_{YX},T_{ZX}}$ for any input-symmetric DBC
has
\begin{equation}
\sum_{j=1}^{l_s} G_{j} = \boldsymbol{1}\boldsymbol{1}^T.
\end{equation}
\end{conjecture}

\section{Discrete Multiplication Degraded Broadcast Channels}
\label{sec:multiplication}

\begin{definition}
{\em (Discrete Multiplication)} A commutative  operation on two inputs from the set
$\{0,1,\cdots,n\}$ is a discrete multiplication if it satisfies the
group axioms on $\{1,\cdots,n\}$, and also produces zero if either input is zero. Use $\otimes$ to denote discrete multiplication.
\end{definition}

\begin{figure}
  \centering
  \includegraphics[width=0.60\textwidth]{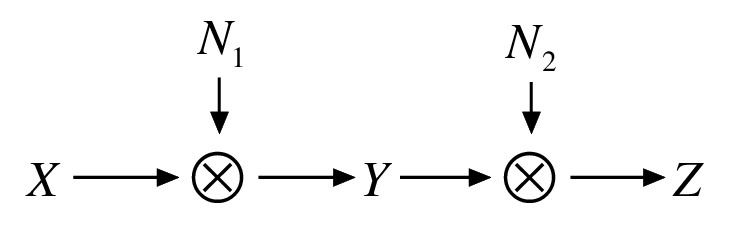}
  \caption{\replaced[BX]{The discrete multiplication degraded broadcast channel.}{The discrete degraded broadcast multiplication channel.}}\label{fig:multiplicationchannel}
\end{figure}

\begin{definition}
{\em (Discrete Multiplication Degraded Broadcast Channel)} A
discrete DBC $X \rightarrow Y \rightarrow Z$ with
$\mathcal{X,Y,Z}$$= \{0,1,\cdots, n\}$ is a discrete multiplication
DBC if there exist two $(n+1)$-ary random variables $N_1$ and $N_2$
such that $Y \sim X \otimes N_1$ and $Z \sim Y \otimes N_2$ as shown
in Figure~\ref{fig:multiplicationchannel}.
\end{definition}

\begin{figure}
  \centering
  \includegraphics[width=0.40\textwidth]{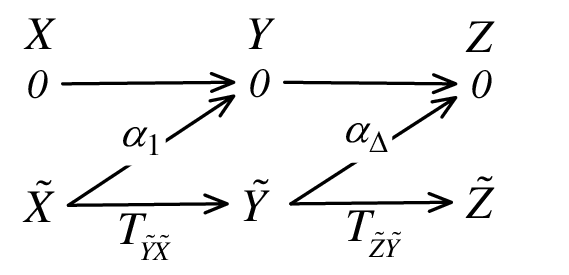}
  \caption{The channel structure of a DBC with erasures.}\label{fig:DBCerasure}
\end{figure}

As an example, the discrete multiplication DBC with $n=1$ is the
broadcast Z channel, which is studied in Section \ref{sec:BZC}. By
the definition of discrete multiplication, the discrete
\added[BX]{multiplication} DBC $X \rightarrow Y \rightarrow Z$ has
the channel structure as shown in Figure~\ref{fig:DBCerasure}. The
sub-channel $\tilde{X} \rightarrow \tilde{Y} \rightarrow \tilde{Z}$
is a group-\replaced{operation}{additive} DBC with transition
matrices $T_{\tilde{Y}\tilde{X}}$ and $T_{\tilde{Z}\tilde{X}} =
T_{\tilde{Z}\tilde{Y}}T_{\tilde{Y}\tilde{X}}$, where
$\tilde{\mbox{$\mathcal{X}$}}$, $\tilde{\mbox{$\mathcal{Y}$}}$,
$\tilde{\mbox{$\mathcal{Z}$}} = \{1,\cdots,n\}$. For the discrete
multiplication DBC $X \rightarrow Y \rightarrow Z$, if the channel
input $X$ is zero, the channel outputs $Y$ and $Z$ are also zeros.
If the channel input is a non-zero symbol, the channel output $Y$ is
zero with probability $\alpha_1$ and $Z$ is zero with probability
$\alpha_2$, where $\alpha_2 = \alpha_1 + (1- \alpha_1)
\alpha_{\Delta}$. Therefore, the transition matrices for $X
\rightarrow Y \rightarrow Z$ are
\begin{equation}
T_{YX}=\begin{bmatrix}
1 & \alpha_1 \boldsymbol{1}^T \\
\boldsymbol{0} & (1-\alpha_1)T_{\tilde{Y}\tilde{X}}
\end{bmatrix},
T_{ZY}=\begin{bmatrix}
1 & \alpha_{\Delta} \boldsymbol{1}^T \\
\boldsymbol{0} & (1-\alpha_{\Delta})T_{\tilde{Z}\tilde{Y}}
\end{bmatrix},
\end{equation}
and
\begin{equation}
T_{ZX}= T_{ZY}T_{YX} =
\begin{bmatrix}
1 & \alpha_{\Delta} \boldsymbol{1}^T \\
\boldsymbol{0} & (1-\alpha_{\Delta})T_{\tilde{Z}\tilde{Y}}
\end{bmatrix}\begin{bmatrix}
1 & \alpha_1 \boldsymbol{1}^T \\
\boldsymbol{0} & (1-\alpha_1)T_{\tilde{Y}\tilde{X}}
\end{bmatrix} = \begin{bmatrix}
1 & \alpha_2 \boldsymbol{1}^T\\
\boldsymbol{0} & (1-\alpha_2)T_{\tilde{Z}\tilde{X}}
\end{bmatrix},
\end{equation}
where \textbf{1} is an all-ones vector and \textbf{0} is an
all-zeros vector.

\subsection{Optimal input distribution}

The sub-channel $\tilde{X} \rightarrow \tilde{Y} \rightarrow
\tilde{Z}$ is a group-\replaced{operation}{additive} DBC, and hence,
$\mbox{$\mathcal{G}$}_{T_{\tilde{Y}\tilde{X}},T_{\tilde{Z}\tilde{X}}}$
is transitive. For any $n \times n$ permutation matrix $\tilde{G}
\in
\mbox{$\mathcal{G}$}_{T_{\tilde{Y}\tilde{X}},T_{\tilde{Z}\tilde{X}}}$
with
$T_{\tilde{Y}\tilde{X}}\tilde{G}=\tilde{\Pi}_{\tilde{Y}\tilde{X}}
T_{\tilde{Y}\tilde{X}}$ and
$T_{\tilde{Z}\tilde{X}}\tilde{G}=\tilde{\Pi}_{\tilde{Z}\tilde{X}}
T_{\tilde{Z}\tilde{X}}$, the $(n+1) \times (n+1)$ permutation matrix
\begin{equation}
G=\begin{bmatrix}
1 & \boldsymbol{0}^{T} \\
\boldsymbol{0} & \tilde{G}
\end{bmatrix}
\end{equation}
has
\begin{equation}
T_{YX}G=\begin{bmatrix}
1 & \alpha_1 \boldsymbol{1}^T \\
\boldsymbol{0} & (1-\alpha_1)T_{\tilde{Y}\tilde{X}}
\end{bmatrix}\begin{bmatrix}
1 & \boldsymbol{0}^{T} \\
\boldsymbol{0} & \tilde{G}
\end{bmatrix}=\begin{bmatrix}
1 & \boldsymbol{0}^{T} \\
\boldsymbol{0} & \tilde{\Pi}_{\tilde{Y}\tilde{X}}
\end{bmatrix}T_{YX},
\end{equation}
and so $G \in \mbox{$\mathcal{G}$}_{T_{YX}}$. Similarly, $G \in
\mbox{$\mathcal{G}$}_{T_{ZX}}$, and hence $G \in
\mbox{$\mathcal{G}$}_{T_{YX},T_{ZX}}$. Therefore, for any $i,j \in
\{1,\cdots,n\}$, there exists a permutation matrix $G \in
\mbox{$\mathcal{G}$}_{T_{YX},T_{ZX}}$ which maps the $(i+1)$-th row
(corresponding to the element $i$) to the $(j+1)$-th row
(corresponding to the element $j$). However, there is no matrix in
$\mbox{$\mathcal{G}$}_{T_{YX},T_{ZX}}$ which maps the first row
(corresponding to the element 0) to other rows (corresponding
non-zero elements) or vice versa. Hence, any permutation matrix $G
\in \mbox{$\mathcal{G}$}_{T_{YX},T_{ZX}}$ has
\begin{equation}
G=\begin{bmatrix} 1 & \boldsymbol{0}^T\\
\boldsymbol{0} & \tilde{G} \end{bmatrix},
\end{equation}
for some $\tilde{G} \in
\mbox{$\mathcal{G}$}_{T_{\tilde{Y}\tilde{X}},T_{\tilde{Z}\tilde{X}}}$.
These results may be summarized in the following lemma:

\begin{lemma}\label{theorem:Multiplication_G}
Let
$\mbox{$\mathcal{G}$}_{T_{\tilde{Y}\tilde{X}},T_{\tilde{Z}\tilde{X}}}
= \{\tilde{G}_1, \cdots , \tilde{G}_l\}$. Hence,
$\mbox{$\mathcal{G}$}_{T_{YX},T_{ZX}} = \{ G_1, \cdots , G_l\}$,
where
\begin{equation}
G_j=\begin{bmatrix}
1 & \boldsymbol{0}^{T} \\
\boldsymbol{0} & \tilde{G}_j
\end{bmatrix},
\end{equation}
for $j=1,\dots, l$.
\end{lemma}

Lemma \ref{theorem:Multiplication_Cstar_Cu} states  that the uniformly distributed $\tilde{X}$ is
optimal for the discrete multiplication DBC.

\begin{lemma}\label{theorem:Multiplication_Cstar_Cu}
Let $\boldsymbol{p}_{X} = [1-q ,q \boldsymbol{p}_{\tilde{X}}^T]^T
\in \Delta_{n+1}$ be the distribution of channel input $X$, where
$\boldsymbol{p}_{\tilde{X}}$ is the distribution of $\tilde{X}$. For
any discrete multiplication DBC,
$\mbox{$\mathcal{C}$}^*_{\boldsymbol{p}_{X}} \subseteq
\mbox{$\mathcal{C}$}^*_{[1-q,q\boldsymbol{u}^T]^T}$ and
$\mbox{$\mathcal{C}$}^{*} = \bigcup_{q \in
[0,1]}\mbox{$\mathcal{C}$}^*_{[1-q,q\boldsymbol{u}^T]^T}$, where
$\boldsymbol{u} \in \Delta_{n}$ denotes the uniform distribution.
\end{lemma}

The proof of Lemma \ref{theorem:Multiplication_Cstar_Cu} is similar
to that of Lemma \ref{theorem:Cstar_Cu} and the details are given in
Appendix \ref{app:C}.

\begin{theorem} \label{theorem:Multiplication_unif_symm}
The capacity region of the discrete multiplication DBC can be
achieved by using transmission strategies where $\tilde{X}$ is
uniformly distributed, i.e., the distribution of $X$ has
$\boldsymbol{p}_{X} = [1-q,q \boldsymbol{u}^T]^T$ for some $q \in
[0,1]$. As a consequence, the capacity region is
\begin{align}\label{eq:multi_capa_symm}
\bar{\text{co}} \Big [ \bigcup_{q \in [0,1]}  \big \{ (R_1,R_2): &
R_1 \leq s - q
h_{n}(T_{\tilde{Y}\tilde{X}}\boldsymbol{e_1}), \nonumber \\
& R_2 \leq h((1-\alpha_2)q) + (1-\alpha_2)q \ln (n) -
F^*_{T_{YX},T_{ZX}}([1-q,q \boldsymbol{u}^T]^T,s) \big\} \Big ].
\end{align}
\end{theorem}

\begin{proof} Let $\boldsymbol{p}_{X} =
[1-q,q\boldsymbol{p}_{\tilde{X}}]^T$ be the distribution of the
channel input $X$, where $\boldsymbol{p}_{\tilde{X}} = [p_1 ,\cdots,
p_n]^T$. Since $\mbox{$\mathcal{G}$}_{T_{\tilde{Y}\tilde{X}}}$ is
transitive and the columns of $T_{\tilde{Y}\tilde{X}}$ are
permutations of each other.
\begin{align}
H(Y|X) & = \sum_{i=0}^{n} \text{Pr}(X=i)H(Y|X=i)\\
 & = (1-q)H(Y|X=0) + \sum_{i=1}^{n} q p_i h_{n}(T_{\tilde{Y}\tilde{X}} \boldsymbol{e_i})\\
 & = \sum_{i=1}^{n} q p_i h_{n}(T_{\tilde{Y}\tilde{X}} \boldsymbol{e_1})\\
 & = q h_{n}(T_{\tilde{Y}\tilde{X}} \boldsymbol{e_1}) \label{eq:multi_unif_symm_1},
\end{align}
which is independent of $\boldsymbol{p}_{X}$. Let
$\mbox{$\mathcal{G}$}_{T_{YX},T_{ZX}} = \{G_1,\cdots,G_l\}$.
\begin{align}
H(Z) & = h_{n+1}(T_{ZX}\boldsymbol{p}_{X}) \\
& = \frac{1}{l}\sum_{i=1}^{l} h_{n+1}\left(T_{ZX}G_{i}\boldsymbol{p}_{X}\right) \\
& \leq h_{n+1}\left(T_{ZX} \frac{1}{l}\sum_{i=1}^{l}G_{i}\boldsymbol{p}_{X}\right) \label{eq:multi_unif_symm_11}\\
& = h_{n+1} \left (T_{ZX} [1-q,q \boldsymbol{u}^T]^T \right) \\
& = h_{n+1} \left ([1-q+\alpha_2 q, (1-\alpha_2)q\boldsymbol{u}]^{T} \right)\\
& = h((1-\alpha_2)q) + (1- \alpha_2)q \ln (n),
\label{eq:multi_unif_symm_2}
\end{align}
where (\ref{eq:multi_unif_symm_11}) follows from Jensen's
inequality \added{and \eqref{eq:multi_unif_symm_2} follows from the grouping rule for entropy \cite[Problem 2.27]{BookCover}}.  By Lemma \ref{theorem:Multiplication_Cstar_Cu},
$\mbox{$\mathcal{C}$}^*_{\boldsymbol{p}_{X}} \subseteq
\mbox{$\mathcal{C}$}^*_{[1-q,q\boldsymbol{u}^T]^T}$ for the discrete
multiplication DBC. Hence,
\begin{equation}
F^*(\boldsymbol{p}_{X},s) \geq F^*([1-q, q\boldsymbol{u}^T]^T,s).
\label{eq:multi_unif_symm_3}
\end{equation}
Plugging (\ref{eq:multi_unif_symm_1}), (\ref{eq:multi_unif_symm_2})
and (\ref{eq:multi_unif_symm_3}) into (\ref{eq:MyCapa3}),  the
capacity region for discrete multiplication DBCs is
\begin{align}
 \bar{\text{co}} \Big [  & \bigcup _{\boldsymbol{p}_{X} \in
\Delta_{k}} \big \{ (R_1  ,R_2):R_1 \leq s - H(Y|X), \nonumber\\
& \qquad R_2  \leq H(Z)- F^*_{T_{YX},T_{ZX}}(\boldsymbol{p}_{X},s)
\big\} \Big ] \label{eq:multi_optinput_1}\\
 \subseteq \;  \bar{\text{co}}  \Big [ & \bigcup _{\boldsymbol{p}_{X} \in
\Delta_{k}} \big \{ (R_1 ,R_2):R_1 \leq s -
h_{n}(T_{\tilde{Y}\tilde{X}} \boldsymbol{e_1}), \nonumber \\
&  \qquad R_2  \leq h((1-\alpha_2)q) + (1- \alpha_2)q \ln (n) \nonumber \\
& \qquad \qquad - F^*_{T_{YX},T_{ZX}}([1-q, q\boldsymbol{u}^T]^T,s)
\big\} \Big ] \label{eq:multi_optinput_2} \\
 = \;  \bar{\text{co}}  \Big [ & \bigcup_{q \in [0,1]}  \big \{ (R_1 ,R_2):
R_1 \leq s - q
h_{n}(T_{\tilde{Y}\tilde{X}}\boldsymbol{e_1}), \nonumber \\
& \qquad  R_2 \leq h((1-\alpha_2)q) + (1-\alpha_2)q \ln (n) \nonumber\\
 & \qquad \qquad - F^*_{T_{YX},T_{ZX}}([1-q,q \boldsymbol{u}^T]^T,s) \big\}
\Big ] \label{eq:multi_optinput_3} \\
=  \bar{\text{co}}  \Big [ & \bigcup
 _{ \boldsymbol{p}_{X} = [1-q,q \boldsymbol{u}^T]^T } \big \{ (R_1 ,R_2):  R_1 \leq s - H(Y|X), \nonumber\\
& \qquad  R_2   \leq H(Z)- F^*_{T_{YX},T_{ZX}}(\boldsymbol{p}_{X},s)
\big\} \Big ] \label{eq:multi_optinput_4} \\
\subseteq \bar{\text{co}} \Big [  & \bigcup _{\boldsymbol{p}_{X} \in
\Delta_{k}} \big \{ (R_1  ,R_2):R_1 \leq s - H(Y|X), \nonumber\\
&  \qquad R_2  \leq H(Z)- F^*_{T_{YX},T_{ZX}}(\boldsymbol{p}_{X},s)
\big\} \Big ], \label{eq:multi_optinput_5}
\end{align}
where $\bar{\text{co}}$ denotes the convex hull of the closure. Note
that (\ref{eq:multi_optinput_1}) and (\ref{eq:multi_optinput_5}) are
identical expressions, hence (\ref{eq:multi_optinput_1} -
\ref{eq:multi_optinput_5}) are all equal. Therefore,
(\ref{eq:multi_optinput_3}) expresses the capacity region for the
discrete multiplication DBC, which also means that the capacity
region can be achieved by using transmission strategies where the
broadcast signal $X$ has distribution $\boldsymbol{p}_{X} =
[1-q,q\boldsymbol{u}^T]^T$ for some $q \in [0,1]$.
\end{proof}

\subsection{Optimality of the NE scheme for DM-DBCs}

\begin{figure}
  \centering
  \includegraphics[width=0.80\textwidth]{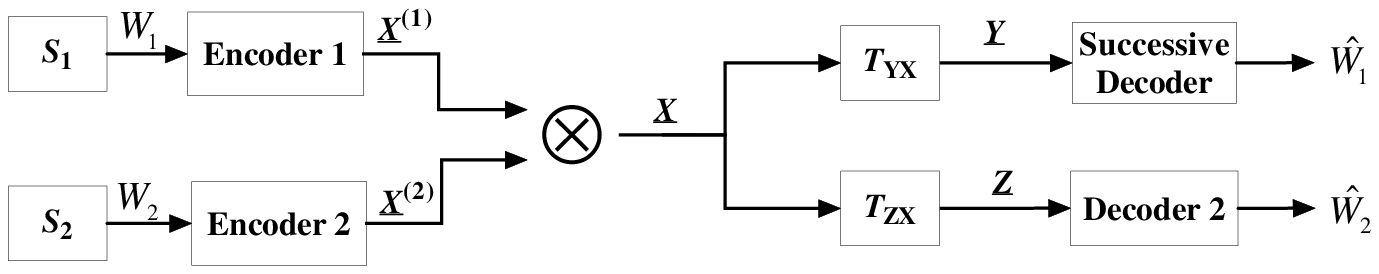}
  \caption{The block diagram of the NE scheme for the discrete multiplication DBC.}\label{fig:multiapproach}
\end{figure}

The NE scheme for the discrete multiplication DBC is shown in
Figure~\ref{fig:multiapproach}. $W_1$ is the message for Receiver 1
who sees the less-degraded channel $T_{YX}$ and $W_2$ is the message
for Receiver 2 who sees the more-degraded channel $T_{ZX}$. The NE
scheme is first to independently encode these two messages into two
codewords $\boldsymbol{X}^{(1)}$ and $\boldsymbol{X}^{(2)}$
respectively where $\mbox{$\mathcal{X}$}^{(1)},
\mbox{$\mathcal{X}$}^{(2)} = \{0,1,\cdots,n\}$, and then to
broadcast $\boldsymbol{X}$ which is obtained by applying the
single-letter function $X = X^{(2)} \otimes X^{(1)}$ on symbols of
codewords $\boldsymbol{X}^{(1)}$ and $\boldsymbol{X}^{(2)}$. The
distribution of $X^{(2)}$ is constrained to be
$\boldsymbol{p}_{X^{(2)}} = [1-q,q\boldsymbol{u}^T]^T$ for some $q
\in [0,1]$ and hence the distribution of the broadcast signal $X$
also has $\boldsymbol{p}_{X} = [1-q,q\boldsymbol{u}^T]^T$ for some
$q \in [0,1]$, which was proved to be the optimal input distribution
for the discrete multiplication DBC. Receiver 2 receives
$\boldsymbol{Z}$ and decodes the desired message directly. Receiver
1 receives $\boldsymbol{Y}$ and successively decodes the message for
Receiver 2 and then for Receiver 1.

Let $\boldsymbol{p}_{X} = [1-q, q \boldsymbol{p}_{\tilde{X}}]^T$ be
the distribution of the channel input $X$, where
$\boldsymbol{p}_{\tilde{X}}$ is the distribution of sub-channel
input $\tilde{X}$. For the discrete multiplication DBC $X
\rightarrow Y \rightarrow Z$, the $\phi$ function is
\begin{align}
\phi (\boldsymbol{p}_{X}, \lambda) & =
h_{n+1}(T_{ZX}\boldsymbol{p}_{X}) - \lambda h_{n+1}(T_{YX}
\boldsymbol{p}_{X})
\\
& = h_{n+1} \left(\begin{bmatrix}
1-q + q \alpha_2 \\
q(1-\alpha_2) T_{\tilde{Z}\tilde{X}}\boldsymbol{p}_{\tilde{X}}
\end{bmatrix} \right ) - \lambda h_{n+1} \left(\begin{bmatrix}
1- q + q \alpha_1 \\
 q (1-\alpha_1) T_{\tilde{Y}\tilde{X}}\boldsymbol{p}_{\tilde{X}}
\end{bmatrix} \right ) \\
& = h(q(1-\alpha_2)) -  q (1-\alpha_2)h_{n}
\left(T_{\tilde{Z}\tilde{X}}\boldsymbol{p}_{\tilde{X}}\right) -
\lambda \left( h( q (1-\alpha_1)) -
q (1-\alpha_1)h_{n} \left(T_{\tilde{Y}\tilde{X}}\boldsymbol{p}_{\tilde{X}}\right)\right) \\
& = h(q \beta_2) - \lambda h(q \beta_1) + q \beta_2 \left (
h_{n}\left(T_{\tilde{Z}\tilde{X}}\boldsymbol{p}_{\tilde{X}}\right) -
\frac{\lambda}{1-\alpha_{\Delta}}
h_{n}\left(T_{\tilde{Y}\tilde{X}}\boldsymbol{p}_{\tilde{X}}\right) \right ) \\
& = h(q \beta_2) - \lambda h(q \beta_1) + q \beta_2
\tilde{\phi}\left(\boldsymbol{p}_{\tilde{X}},\frac{\lambda}{1-
\alpha_{\Delta}}\right), \label{multi_phi}
\end{align}
where $\beta_1 = 1-\alpha_1$, $\beta_2 = 1-\alpha_2$, and
$\tilde{\phi}(\boldsymbol{q}, \lambda) \triangleq
h_{n}(T_{\tilde{Z}\tilde{X}}\boldsymbol{q}) - \lambda
h_{n}(T_{\tilde{Y}\tilde{X}}\boldsymbol{q})$ is the $\phi$ function
defined on the \added[BX]{group-operation degraded broadcast}
sub-channel $\tilde{X} \rightarrow \tilde{Y} \rightarrow \tilde{Z}$.

Define $\tilde{\psi}(\boldsymbol{q},\lambda) \triangleq
\underline{\text{env}}_{\boldsymbol{q}} \tilde{\phi}(\boldsymbol{q},
\lambda)$ as the $\psi$ function for \added[BX]{group-operation
degraded broadcast} sub-channel $\tilde{X} \rightarrow \tilde{Y}
\rightarrow \tilde{Z}$ where the lower envelope is taken with
respect to $\boldsymbol{q}$.

For the channel $X \rightarrow Y \rightarrow Z$, define the lower
envelope of $\phi (\boldsymbol{p}_{X}, \lambda)$ with respect to
$\boldsymbol{p}_{\tilde{X}}$ (not with respect to $\boldsymbol{p}_{X}$) as
\begin{align}
\varphi(q,\boldsymbol{p}_{\tilde{X}},\lambda ) & \triangleq
\underline{\text{env}}_{\boldsymbol{p}_{\tilde{X}}} \phi
(\boldsymbol{p}_{X}, \lambda) \\
&=h(q \beta_2) - \lambda h(q \beta_1) + q \beta_2
\tilde{\psi}\left(\boldsymbol{p}_{\tilde{X}},\frac{\lambda}{1-
\alpha_{\Delta}}\right). \label{multi_varphi}
\end{align}
Therefore, the $\psi$ function for $X \rightarrow Y \rightarrow Z$
has
\begin{align}
\psi(\boldsymbol{p}_{X}, \lambda) & =
\underline{\text{env}}_{\boldsymbol{p}_{X}} \phi
(\boldsymbol{p}_{X}, \lambda) \\
& = \underline{\text{env}}_{\boldsymbol{p}_{X}}
\varphi(q,\boldsymbol{p}_{\tilde{X}},\lambda).
\end{align}

\begin{lemma}\label{theorem:opt_tildephi}
$\psi([1-q,q\boldsymbol{u}^T]^T, \lambda)$ is the lower
envelope of $\varphi(q,\boldsymbol{u},\lambda )$ with respect to $q$, i.e.,
\begin{equation}
\psi([1-q,q\boldsymbol{u}^T]^T, \lambda) =
\underline{\text{env}}_{q} \varphi(q,\boldsymbol{u},\lambda).
\end{equation}
\end{lemma}

The proof is given in Appendix \ref{app:D}. Lemma \ref{theorem:opt_tildephi}
indicates that the lower envelope of $\phi(\cdot, \lambda)$ with respect to
$\boldsymbol{p}_{X} = [1-q,q \boldsymbol{u}^T]^T$ can be obtained two steps by decomposing $\boldsymbol{p}_{X}$ into $q$ and $\boldsymbol{p}_{\tilde{X}}$.   The first step is for any fixed $q$, the lower envelope of $\phi(\boldsymbol{p}_{X}, \lambda)$ with respect to $\boldsymbol{p}_{\tilde{X}}$
is $\varphi(q,\boldsymbol{p}_{\tilde{X}},\lambda )$. Second, for
$\boldsymbol{p}_{\tilde{X}}= \boldsymbol{u}$, the lower envelope of
$\varphi(q,\boldsymbol{u},\lambda )$ with respect to $q$ coincides with
$\psi(\boldsymbol{p}_{X}, \lambda)$, which is the desired lower envelope of
$\phi(\boldsymbol{p}_{X}, \lambda)$ with respect to $\boldsymbol{p}_{X}$.

Now we state and prove that NE is optimal for the
discrete multiplication DBC.

\begin{theorem}\label{theorem:opt_multiplication}
NE achieves the capacity region for the discrete
multiplication DBC.
\end{theorem}

\begin{figure}
  \centering
  \includegraphics[width=0.60\textwidth]{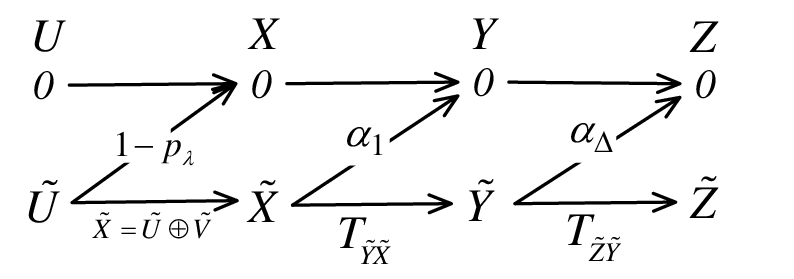}
  \caption{The optimal transmission strategy for the discrete multiplication degraded broadcast channel}\label{fig:multi_opt}
\end{figure}

\begin{proof} This proof shows that combining NE for the broadcast Z channel with NE for the group-operation DBC achieves the capacity region of the discrete multiplication DBC.  This encoding is also the NE for this
channel.

Theorem \ref{theorem:Multiplication_unif_symm} shows that the
capacity region for the discrete multiplication DBC can be achieved
by using transmission strategies with uniformly distributed
$\tilde{X}$, i.e., the input distribution $\boldsymbol{p}_{X} =
[1-q, q \boldsymbol{u}^T]^T$. By Lemma \ref{theorem:opt_tildephi},
for \added{such a} $\boldsymbol{p}_{X}$,
$\psi([1-q,q\boldsymbol{u}^T]^T, \lambda)$ can be attained by the
convex combination of points on the graph of
$\varphi(q,\boldsymbol{u},\lambda )$. \replaced[BX]{Recall}{Note}
that
\begin{align}
\varphi(q,\boldsymbol{u},\lambda ) &= h(q \beta_2) - \lambda h(q
\beta_1) + q \beta_2 \tilde{\psi}\left(\boldsymbol{u},
\frac{\lambda}{1-\alpha_{\Delta}}\right)\\
&= \phi_Z(q,\lambda)+ q \beta_2 \tilde{\psi}\left(\boldsymbol{u},
\frac{\lambda}{1-\alpha_{\Delta}}\right)\, ,
\end{align}
\replaced{where $\phi_Z$ is  $\phi$ for the broadcast Z channel and
$\tilde{\psi}$ is $\psi$ for the group-operation DBC.}{which is the
sum of $\phi(q,\lambda)$ for the broadcast Z channel and $q$ times
the constant $\beta_2 \tilde{\psi}(\boldsymbol{u},
\frac{\lambda}{1-\alpha_{\Delta}})$.} \deleted[BX]{We can separately
optimize $\phi_Z$  and $\tilde{\psi}$ using results derived above
for the broadcast Z channel and the group-operation DBC to identify
the points of $\varphi(q,\boldsymbol{u},\lambda )$  to convexly
combine to produce an optimal strategy for the discrete
multiplication DBC.}

Hence, by a discussion analogous to Section \ref{sec:BZC}, $\psi([1-q,q\boldsymbol{u}^T]^T,
\lambda)$ can be attained by the convex combination of 2 points on
the graph of $\varphi(q,\boldsymbol{u},\lambda )$. One point is at
$q=0$ and $\varphi(0,\boldsymbol{u},\lambda ) = 0$. The other point
is at $q = p_{\lambda}$, determined by solving $\ln
(1-\beta_2 p_{\lambda}) = \lambda \ln (1-\beta_1 p_{\lambda})$ for $p_{\lambda}$.

Note that the point (0,0) on the graph of
$\varphi(q,\boldsymbol{u},\lambda )$ is also on the graph of
$\phi(\boldsymbol{p}_{X},\lambda)$. By Theorem \ref{theorem:Fall},
the point $(p_{\lambda}, \varphi(p_{\lambda},\boldsymbol{u},\lambda
))$ is the convex combination of $n$ points on the graph of
$\phi(\boldsymbol{p}_{X},\lambda)$, which corresponds to the
group-\replaced{operation}{addition} encoding approach for the sub-channel $\tilde{X}
\rightarrow \tilde{Y} \rightarrow \tilde{Z}$ because the
group-\replaced{operation}{addition} encoding approach is the optimal NE scheme for the
group-\replaced{operation}{additive} DBC $\tilde{X} \rightarrow \tilde{Y} \rightarrow
\tilde{Z}$. Therefore, by Theorem \ref{theorem:Fall}, an optimal
transmission strategy for the discrete multiplication DBC $X
\rightarrow Y \rightarrow Z$ is NE as shown in
Figure~\ref{fig:multi_opt}.\end{proof}

If the auxiliary random variable $U$ is 0, then the channel input
$X$ equals \deleted{to} 0 with probability 1. If $U$ is non-zero,
then $X$ equals \deleted{to} 0 with probability $1-p_{\lambda}$. In
the case where $U$ and $X$ are both non-zero, $\tilde{X}$ can be
obtained as $\tilde{X} = \tilde{U} \oplus \tilde{V}$, where $\oplus$
is the group \replaced{operation}{addition} defined in the
group-\replaced{operation}{additive} degraded broadcast sub-channel
$\tilde{X} \rightarrow \tilde{Y} \rightarrow \tilde{Z}$. Here
$\tilde{U}$ is uniformly distributed and $\tilde{V}$ is an $n$-ary
random variable. \added[BX]{In order to achieve a pareto-optimal
rate pair which maximizes $(R_2 + \lambda R_1)$ for the discrete
multiplication DBC $X \rightarrow Y \rightarrow Z$, the crossover
probability $1-p_{\lambda}$ is determined by $\ln (1-\beta_2
p_{\lambda}) = \lambda \ln (1-\beta_1 p_{\lambda})$, and the
distribution of $\tilde{V}$ should be the one which also maximizes
$(\tilde{R}_2 + \frac{\lambda}{1-\alpha_{\Delta}} \tilde{R}_1)$ for
the group-operation DBC $\tilde{X} \rightarrow \tilde{Y} \rightarrow
\tilde{Z}$.}

Since the NE scheme is optimal for discrete multiplication DBCs, its
achievable rate region is the capacity region for discrete
multiplication DBCs. Hence, the capacity region for the discrete
multiplication DBC in Figure~\ref{fig:multiplicationchannel} is
\begin{align}
\bar{\text{co}} \Big[ &
\bigcup_{p_{U}=[1-q,q\boldsymbol{u}^T]^T,p_{V} \in \Delta_{n+1}}
\!\! \big\{ (R_1 ,R_2): R_2
 \leq H(U \!\otimes\! V \!\otimes\! N_2) \!-\! H(U \!\otimes\! V \! \otimes\! N_2 |U) \nonumber \\
R_1 & \!\leq\! H(U \!\otimes\! V \!\otimes\! N_1 | U) \!-\! H(U
\!\otimes\! V \!\otimes\! N_1 | U \!\otimes\! V) \big \} \Big ].
\end{align}

\section{Conclusions}
\label{sec:conclusion}

This paper extends the set of degraded broadcast channels for which
relatively simple encoding schemes are known to achieve capacity.
These results are obtained by extending the input symmetry and
conditional entropy bound concepts of Wyner and Witsenhausen to
degraded broadcast channels.  This paper introduces permutation
encoding as a relatively simple capacity-achieving approach for
input-symmetric degraded broadcast channels.  This paper also
introduces the concept of natural encoding and shows that natural
encoding achieves the boundary of the capacity region for the
broadcast Z channel with any number of receivers, for the
two-receiver group-\replaced{operation}{additive} degraded broadcast
channel, and (by combining the two previous results) the
two-receiver discrete multiplication degraded broadcast channel.

The capacity-region characterization approach that we use has the
potential to provide explicit characterizations of degraded
broadcast channel capacity regions.  As examples we provide explicit
capacity regions for the two-\replaced[BX]{receiver}{user}
binary-symmetric degraded broadcast channel and the
two-\replaced[BX]{receiver}{user} broadcast Z channel.

A main result \replaced{of}{if} this paper is that simple approaches such as natural encoding and permutation encoding  achieve the capacity
region of degraded broadcast channels much more often that has been previously known.   It would seem that there are more such cases where natural encoding achieves the DBC capacity region waiting to be identified. It remains an open problem to prove a general theorem establishing the optimality of natural encoding over a suitably large class of DBCs. The results of this paper also open interesting problems in channel coding to find practical channel codes that use permutation encoding or natural encoding to approach the channel capacity region for the degraded broadcast channels studied in this paper.

\comment{This paper exploresrelatively simple optimal encoders for the
degraded broadcast channel. By extending the input-symmetry and
conditional entropy bound ideas of Wyner and Witsenhausen, this
paper introduced and proved the optimality of the permutation
encoding approach for input-symmetric DBCs and showed that the
natural encoding scheme can achieve the boundary of the capacity
region for the multi-receiver broadcast Z channel, the two-receiver
group-additive DBC, and (by combining the previous two results) the
two-receiver discrete multiplication DBC. Along the way, this paper
has provided closed-form expressions for the capacity regions of
those DBCs. In conclusion, natural encoding achieves the capacity
region of DBCs much more often that has been previously known.  In
fact, it would seem that there are more such cases where natural
encoding achieves the DBC capacity region waiting to be identified.
It remains an open problem to prove a general theorem establishing
the optimality of natural encoding over a suitably large class of
DBCs. The results of this paper also open interesting problems in
channel coding to find practical channel codes for the DBCs examined
in this paper.}

\section{Acknowledgment}
\label{sec:thank}

The authors wish to thank and acknowledge Emre Telatar for his helpful comments, which were essential to properly framing our natural encoding results within the context of existing independent encoding approaches.  Also, the authors are gratefully indebted to the Associate Editor Elza Erkip and to the reviewers.  Their careful reading of the manuscript and many detailed and helpful comments greatly improved the final paper.


\appendices
\section{A Simple Independent Encoding Scheme}\label{app:A}
This appendix presents a simple independent encoding scheme made
known to us by Telatar \cite{TelatarPrivate} which achieves the
capacity region for DBCs. The scheme generalizes to any number of
receivers, but showing the two-receiver case suffices to explain the
approach. It indicates that any achievable rate pair $(R_1,R_2)$ for
a DBC can be achieved by combining symbols from independent encoders
with a single-letter function. The independent encoders operate
using two codebooks $\{v^n(i):i=1,\cdots,2^{nR_1}\}$,
$\{u^n(j):j=1,\cdots,2^{nR_2}\}$ and a single-letter function
$f(v,u)$. In order to transmit the message pair $(i,j)$, the
transmitter sends the sequence
$f(v_1(i),u_1(j)),\cdots,f(v_n(i),u_n(j))$. The scheme is described
below:

\begin{lemma}\label{theorem:UVfConstruction}
Suppose $U$ and $X$ are discrete random variables with joint
distribution $p_{U,X}(u,x)$. There exists a random vector $V$ independent of $U$ and
a deterministic function $f$ such that the pair $(U, f(V,U))$ has
joint distribution $p_{U,X}(u,x)$. \cite{TelatarPrivate}
\end{lemma}

\begin{proof} Suppose $U$ and $X$ take values in $\{1, \cdots, l\}$
and $\{1, \cdots, k\}$ respectively. Let $V = (V_1, \cdots, V_l)$,
independent of $U$, be a random variable taking values in $\{1,
\cdots, k\}^l$ with $\text{Pr}(V_j=i)=p_{X|U}(i|j)$. Set
$f((v_1,\cdots,v_l),u) = v_u$. Then we have
\begin{align}
\text{Pr}(U=u, f(V,U) = x) & = \text{Pr}(U=u, V_u=x) \nonumber \\
& = \text{Pr}(U=u)Pr(V_u=x)\nonumber\\
& = p_U(u)p_{X|U}(x|u) \nonumber \\
& = p_{U,X}(u,x).
\end{align}
\end{proof}

If the rate pair $(R_1,R_2)$ is achievable for a DBC $X \rightarrow
Y \rightarrow Z$, there exists an auxiliary random variable $U$ such
that
\begin{align}
\text{(a) } & U \rightarrow X \rightarrow Y \rightarrow Z;
\nonumber\\
\text{(b) } & I(X;Y|U) \geq R_1; \nonumber \\
\text{(c) } & I(U;Z) \geq R_2.
\end{align}
Apply Lemma \ref{theorem:UVfConstruction} to find $V$ independent of
$U$ and the deterministic function $f(v,u)$ such that the pair $(U,
f(V,U))$ has the same joint distribution as that of $(U,X)$.
Randomly and independently choose codewords $\left \{
v^n(1),\cdots,v^n(2^{nR_1})\right \}$ according to
$p(v^n)=p_{V}(v_1) \cdots p_{V}(v_n)$, and choose codewords $\left
\{ u^n(1),\cdots,u^n(2^{nR_2}) \right \}$ according to
$p(u^n)=p_{U}(u_1) \cdots p_{U}(u_n)$. To send message pair $(i,j)$,
the encoder transmits $f(v_1(i),u_1(j)),\cdots,f(v_n(i),u_n(j))$.

Using a typical-set-decoding random-coding argument, the weak
decoder, given $z^n$, searches for the unique $j'$ such that
$(z^n,u^n(j'))$ is jointly typical. The error probability converges
to zero as $n$ goes to infinity since $R_2 \leq I(U;Z)$. The strong
decoder, given $y^n$, also searches for the unique $j'$ such that
$(y^n,u^n(j'))$ is jointly typical, and then searches for the unique
$i'$ such that $(y^n,v^n(i'))$ is jointly typical given $u^n(j')$.
The error probability converges to zero as $n$ goes to infinity
since
\begin{equation}
R_2 \leq I(U;Z) \leq I(U;Y),
\end{equation}
and
\begin{align}
R_1 & \leq I(X;Y|U) \nonumber\\
& =H(Y|U)-H(Y|f(V,U),U) \nonumber\\
& =H(Y|U)-H(Y|f(V,U),U,V) \nonumber\\
& =H(Y|U) - H(Y|U,V) \nonumber \\
& =I(V;Y|U).
\end{align}

\section{Proof of (\ref{eq:contradiction})}\label{app:B}
\begin{proof}[Proof of (\ref{eq:contradiction})]
Plugging $j=1$ in
(\ref{eq:capa_KBZC_3}), we have
\begin{equation}
H(\boldsymbol{Y}^{(1)} | W_{2}, \cdots, W_{K}) -
H(\boldsymbol{Y}^{(1)} | W_{1}, \cdots, W_{K}) \geq N
\frac{q}{t_1}h(\beta_{1}t_1) - N q h(\beta_{1}) - o(\epsilon)
\end{equation}
or
\begin{equation}
H(\boldsymbol{Y}^{(1)} | W_{2}, \cdots, W_{K}) \geq N
\frac{q}{t_1}h(\beta_{1}t_1) - o(\epsilon), \label{eq:appendixD1}
\end{equation}
since
\begin{align}
H(\boldsymbol{Y}^{(1)} | W_{1}, \cdots, W_{K}) & =
H(\boldsymbol{Y}^{(1)} |
\boldsymbol{X})  \label{AppendA1} \\
& = \sum_{i=1}^{N} H(Y^{(1)}_{i} | \boldsymbol{X})  \label{AppendA2} \\
& = \sum_{i=1}^{N} H(Y^{(1)}_{i} | X_i)  \label{AppendA3} \\
& = \sum_{i=1}^{N} \textrm{Pr}(X_i = 0) h(\beta_1) \label{AppendA4} \\
& = N q h(\beta_1).
\end{align}
Some of these steps are justified as follows:
\begin{itemize}
\item (\ref{AppendA1}) follows \replaced[BX]{since}{from the fact that} $\boldsymbol{X}$ is a function of $(W_1, \cdots, W_K)$;
\item (\ref{AppendA2}) follows from the conditional independence of
$Y^{(1)}_{i}$,$i=1,\cdots,N$, given $\boldsymbol{X}$;
\item (\ref{AppendA3}) follows from the conditional independence of
$Y^{(1)}_{i}$ and $(X_1,\cdots,X_{i-1},X_{i+1},\cdots, X_{N})$ given
$X_i$.
\end{itemize}
Inequality (\ref{eq:appendixD1}) indicates that
\begin{equation}
H(\boldsymbol{Y}^{(j)}|W_{j+1},\cdots,W_{K}) \geq N
\frac{q}{t_j}h(\beta_{j}t_j) - o(\epsilon), \label{eq:capa_KBZC_4}
\end{equation}
is true for $j=1$. The rest of the proof is by induction. We assume
that (\ref{eq:capa_KBZC_4}) is true for $j$, which means
\begin{align}
H(\boldsymbol{Y}^{(j)}|W_{j+1},\cdots,W_{K}) & \geq N \left [
\frac{q}{t_j}h(\beta_{j}t_j) - \frac{o(\epsilon)}{N} \right ] \label{AppendA8}\\
& = N
\frac{q}{t_j+\frac{\tau(\epsilon)}{N}}h(\beta_{j}(t_j+\frac{\tau(\epsilon)}{N})),
\end{align}
where the function $\tau(\epsilon) \rightarrow 0$ as $\epsilon
\rightarrow 0$, since $\frac{q}{t_j}h(\beta_{j}t_j)$ is continuous
in $t_j$. Applying Lemma \ref{theorem:vectorZ} to the Markov chain
$(W_{j+1},\cdots,W_{K}) \rightarrow \boldsymbol{X} \rightarrow
\boldsymbol{Y}^{(j)} \rightarrow \boldsymbol{Y}^{(j+1)}$, we have
\begin{align}
H(\boldsymbol{Y}^{(j+1)}| W_{j+1},\cdots,W_{K}) & \geq N
\frac{q}{t_j+\frac{\tau(\epsilon)}{N}}h(\beta_{j+1}(t_j+\frac{\tau(\epsilon)}{N}))\\
& = N
 \frac{q}{t_j} h(\beta_{j+1}t_j) + o(\epsilon). \label{AppendA5}
\end{align}
Considering (\ref{eq:capa_KBZC_3}) for $j+1$, we have
\begin{equation}
H(\boldsymbol{Y}^{(j+1)} | W_{j+2}, \cdots, W_{K}) -
H(\boldsymbol{Y}^{(j+1)} | W_{j+1}, \cdots, W_{K}) \geq N
\frac{q}{t_{j+1}}h(\beta_{j+1}t_{j+1}) - N
\frac{q}{t_{j}}h(\beta_{j+1}t_{j}) - o(\epsilon). \label{AppendA6}
\end{equation}
Substitution of (\ref{AppendA5}) in  (\ref{AppendA6}) yields
\begin{equation}
H(\boldsymbol{Y}^{(j+1)} | W_{j+2}, \cdots, W_{K}) \geq N
\frac{q}{t_{j+1}}h(\beta_{j+1}t_{j+1}) - o(\epsilon),
\label{AppendA7}
\end{equation}
which establishes the induction. Finally, for $j \geq d$, $N \delta$
should be added to the right side of (\ref{AppendA8}) because of the
presence of $\delta$ in (\ref{eq:rates}) for $j = d$, and hence, of
$N \delta$ in (\ref{eq:capa_KBZC_3}).
\end{proof}

\section{Proof of Lemma \ref{theorem:Multiplication_Cstar_Cu}} \label{app:C}
\begin{proof}[Proof of Lemma \ref{theorem:Multiplication_Cstar_Cu}]
Let
$\mbox{$\mathcal{G}$}_{T_{YX},T_{ZX}} = \{G_1,\cdots,G_l\}$. For any
$(s, \eta) \in \mbox{$\mathcal{C}$}^*_{\boldsymbol{p}_{X}}$, where
$\boldsymbol{p}_{X} = [1-q, q \boldsymbol{p}_{\tilde{X}}^T]^T$, one
has $(\boldsymbol{p}_{X}, s, \eta) \in \mbox{$\mathcal{C}$}$. Since
Lemma \ref{theorem:p2Gp} and Corollary \ref{theorem:Cp2CGp} also
hold for the discrete multiplication DBC, $( G_j
\boldsymbol{p}_{X},s,\eta) \in \mbox{$\mathcal{C}$}$ for all
$j=1,\cdots,l$.  By the convexity of the set $\mathcal{C}$,
\begin{equation}
(\boldsymbol{q},s,\eta) = \left(  \sum_{j=1}^{l}
 \frac{1}{l}G_j \boldsymbol{p}_{X} \; , s , \eta \right) \in \mathcal{C},
\end{equation}
where $\boldsymbol{q} = \sum_{j=1}^l \frac{1}{l}G_j
\boldsymbol{p}_{X}$. Since $\mbox{$\mathcal{G}$}_{T_{YX},T_{ZX}}$ is
a group, for any permutation matrix $G' \in
\mbox{$\mathcal{G}$}_{T_{YX},T_{ZX}}$,
\begin{equation}
G' \boldsymbol{q} \; = \! \sum_{j=1}^l \!\! \frac{1}{l}G' G_j
\boldsymbol{p}_{X} \; = \! \sum_{j=1}^l \!\! \frac{1}{l}G_j
\boldsymbol{p}_{X} \; =\; \boldsymbol{q}.
\end{equation}
Hence, the $(i+1)$-th entry and the $(j+1)$-th entry of
$\boldsymbol{q}$ are the same if $G'$ permutes the $(i+1)$-th row to
the $(j+1)$-th row for $i,j \in \{1,\cdots,n\}$. Therefore, the
second to the $(n+1)$-th entries of $\boldsymbol{q}$ are all the
same because the set $\mbox{$\mathcal{G}$}_{T_{YX},T_{ZX}}$ for the
discrete multiplication DBC permutes the $(i+1)$-th row to the
$(j+1)$-th row for all $i,j \in \{1,\cdots,n\}$. Furthermore, no
matrix in $\mbox{$\mathcal{G}$}_{T_{YX},T_{ZX}}$ maps the first row
 to other rows, hence the first entry of
$\boldsymbol{q}$ is the same as the first entry of
$\boldsymbol{p}_{X}$.  Therefore, $\boldsymbol{q} =
[1-q,q\boldsymbol{u}^T]^T$. This implies that $(s,\eta) \in
\mbox{$\mathcal{C}$}^*_{[1-q,q\boldsymbol{u}^T]^T}$, and hence
$\mbox{$\mathcal{C}$}^*_{\boldsymbol{p}_{X}} \subseteq
\mbox{$\mathcal{C}$}^*_{[1-q,q\boldsymbol{u}^T]^T}$. Therefore,
$\mbox{$\mathcal{C}$}^{*} = \bigcup_{q \in
[0,1]}\mbox{$\mathcal{C}$}^*_{[1-q,q\boldsymbol{u}^T]^T}$.
\end{proof}

\section{Proof of Lemma \ref{theorem:opt_tildephi}} \label{app:D}
\begin{proof}[Proof of Lemma \ref{theorem:opt_tildephi}]
$\psi(\boldsymbol{p}_{X}, \lambda)$ is the lower envelope of
$\varphi(q, \boldsymbol{p}_{\tilde{X}}, \lambda)$ with respect to
$\boldsymbol{p}_{X}$. For $\boldsymbol{p}_{X} = [1-q,
q\boldsymbol{u}^T]^T$, suppose the point $(\boldsymbol{p}_{X},
\psi(\boldsymbol{p}_{X}, \lambda))$ is the convex combination of
$n+1$ points $((q_i,\boldsymbol{t}_i), \varphi(q_i,
\boldsymbol{t}_i, \lambda))$ on the graph of $\varphi(q,
\boldsymbol{p}_{\tilde{X}}, \lambda)$ with weights $w_i$ for
$i=1,\cdots, n+1$. Therefore,
\begin{equation}
q = \sum_{i=1}^{n+1} w_iq_i,
\end{equation}
\begin{equation}
\boldsymbol{u} = \sum_{i=1}^{n+1} w_i\boldsymbol{t}_i,
\end{equation}
\begin{equation}
\psi(\boldsymbol{p}_{X}, \lambda) = \sum_{i=1}^{n+1} w_i
\varphi(q_i, \boldsymbol{t}_i, \lambda).
\end{equation}
By Lemma \ref{theorem:Cstar_Cu}, for the group-\replaced{operation}{additive} degraded
broadcast sub-channel, one has
$\mbox{$\mathcal{C}$}^{*}_{\boldsymbol{t}} \subseteq
\mbox{$\mathcal{C}$}^*_{\boldsymbol{u}}$ for any $\boldsymbol{t}$.
Hence, from (\ref{eq:intercept2}),
$\tilde{\psi}(\boldsymbol{t}, \lambda) \geq
\tilde{\psi}(\boldsymbol{u}, \lambda)$ for any $\boldsymbol{t}$,
and so
\begin{equation}
\varphi(q_i,\boldsymbol{t}_i,\lambda) \geq
\varphi(q_i,\boldsymbol{u},\lambda).
\end{equation}
Therefore, the convex combination of $n+1$ points
$((q_i,\boldsymbol{u}), \varphi(q_i, \boldsymbol{u}, \lambda))$ with
weights $w_i$ has
\begin{equation}
\sum_{i=1}^{n+1} w_iq_i = q,
\end{equation}
and
\begin{equation}
 \sum_{i=1}^{n+1} w_i
\varphi(q_i, \boldsymbol{u}, \lambda) \leq \sum_{i=1}^{n+1} w_i
\varphi(q_i, \boldsymbol{t}_i, \lambda) =\psi(\boldsymbol{p}_{X},
\lambda).
\end{equation}
On the other hand, since $\psi(\boldsymbol{p}_{X}, \lambda)$ is the
lower envelope of $\varphi(q, \boldsymbol{p}_{\tilde{X}}, \lambda)$
with respect to $\boldsymbol{p}_{X}$, $\sum_{i=1}^{n+1} w_i \varphi(q_i,
\boldsymbol{u}, \lambda) \geq \psi(\boldsymbol{p}_{X}, \lambda)$ and
hence $\sum_{i=1}^{n+1} w_i \varphi(q_i, \boldsymbol{u}, \lambda) =
\psi(\boldsymbol{p}_{X}, \lambda)$. Therefore,
$\psi([1-q,q\boldsymbol{u}^T]^T, \lambda)$ can be attained as the
convex combination of points on the graph of $\varphi(q,
\boldsymbol{u},\lambda)$ only in the dimension of $q$.
\end{proof}
\bibliographystyle{unsrt}
{\bibliography{broadcastChan}}
\newpage

\end{document}